\documentclass[12pt]{article}
\usepackage{authblk}
\usepackage{geometry}
\usepackage{amsmath}
\usepackage{amssymb}
\usepackage{amsthm}
\usepackage{braket}
\usepackage{mathrsfs}
\usepackage{subfigure}
\usepackage{customcommands}
\usepackage{Nettastyle}
\newcommand{\tr}{\operatorname{tr}}
\DeclareSymbolFont{matha}{OML}{txmi}{m}{it}
\DeclareMathSymbol{v}{\mathord}{matha}{118}
\title{Coarse Graining Holographic Black Holes
}
\author[1]{Netta Engelhardt}
\author[2]{and Aron C. Wall}

\affiliation[1]{Department of Physics, Princeton University, Princeton, NJ 08544, USA}
\affiliation[2]{Stanford Institute for Theoretical Physics, 382 Via Pueblo, Stanford University, Stanford, CA, 94305}

\emailAdd{nengelhardt@princeton.edu}
\emailAdd{aroncwall@gmail.com}

\abstract{We expand our recent work on the \textit{outer entropy}, a holographic coarse-grained entropy defined by maximizing the boundary entropy while fixing the classical bulk data outside some surface.  When the surface is marginally trapped and satisfies certain ``minimar'' conditions, we prove that the outer entropy is exactly equal to a quarter the area (while for other classes of surfaces, the area gives an upper or lower bound).  We explicitly construct the entropy-maximizing interior of a minimar surface, and show that it satisfies the appropriate junction conditions.  This provides a statistical explanation for the area-increase law for spacelike holographic screens foliated by minimar surfaces.  Our construction also provides an interpretation of the area for a class of non-minimal extremal surfaces.

On the boundary side, we define an increasing \textit{simple entropy} by maximizing the entropy subject to a set of ``simple experiments'' performed after some time.  We show (to all orders in perturbation theory around equilibrium) that the simple entropy is the boundary dual to our bulk construction.}

\begin{document}

\maketitle

\section{Introduction} \label{sec:intro}

One of the primary goals of quantum gravity is a complete description of the black hole interior. This description is being pursued via numerous methods, from the AdS/CFT correspondence~\cite{Mal97, GubKle98, Wit98a}, to black hole microstate counting (see literature starting with~\cite{StrVaf96}), and the generalized holographic principle~\cite{Tho93, Sus95, Bou99d} among others (see~\cite{Pol16} for a review). As an ostensible nonperturbative quantum theory of gravity, AdS/CFT in particular has tremendous potential for shedding light on the physics inside the black hole.

The fine-grained entropy of a holographic black hole is given by the HRT formula \cite{RyuTak06, HubRan07}.  As applied to the case of an eternal black hole (which represents an entangled state of two boundary CFT's \cite{Mal01}), the HRT formula tells us that the von Neumann entropy $S_{vN}$ of either CFT is given by the area of a certain compact ``extremal'' surface $X$ (whose area is stationary under variations) lodged inside the throat, which separates the two boundaries:
\be\label{HRTintro}
S_{vN} = \frac{\text{Area}[X]}{4G\hbar}.
\ee
The HRT entropy is time-independent (in the sense that it is independent of the choice of Cauchy slice), so it does not evolve even if we send matter into the black hole.  And for a classical black hole that forms from collapse, $X$ is given by the empty set so $S_{vN}$ vanishes.  This is because the HRT is a fine-grained quantity, i.e. it does not involve any kind of coarse-graining over the thermalized degrees of freedom.  Hence, it does not allow us to define a nontrivial second law, nor does it allow us to interpret the changing area of a black hole horizon as an entropy.  For this, we need a definition of coarse-grained entropy.

A natural framework of coarse-graining, advocated by Jaynes~\cite{Jay57a, Jay57b}, is to maximize the von Neumann entropy $S_{vN} = -\tr (\rho \ln \rho)$ while holding certain quantities fixed.  In our case, we wish to hold fixed the classical bulk data outside of some surface $\sigma$.  The information outside of $\sigma$ will play the role of the ``macrostate'', i.e. the information that is accessible to an exterior observer.  The information inside of $\sigma$ will play the role of the ``microstate'', i.e. the forgotten information which must be coarse-grained over\footnote{Our coarse-grained entropy depends on the choice of the surface $\sigma$. We believe that this is analogous to the ambiguity in thermodynamics, where it is also necessary to devise a prescription for a demarcation between macrostate and microstate, which to some extent is dependent on the scheme.}.  From a bulk perspective, this is justified insofar as an observer living outside of $\sigma$ will find the data in the exterior of $\sigma$ easy to measure, while the data in the interior of $\sigma$ is hard to measure.  Although \emph{in principle}, the data in the interior must be holographically encoded in the boundary CFT, in practice recovery is difficult since the data is encoded in subtle thermalized correlations.  This explains why, at the level of the classical bulk, there is an effective notion of causality in which information appears to be lost once it falls behind a black hole horizon.  As we will see, this notion of causal coarse-graining is holographically dual to a thermodynamic coarse-graining on the boundary.

\begin{figure}\label{fig:innerouterdecomp}
\begin{center}
\includegraphics[width=0.4 \textwidth]{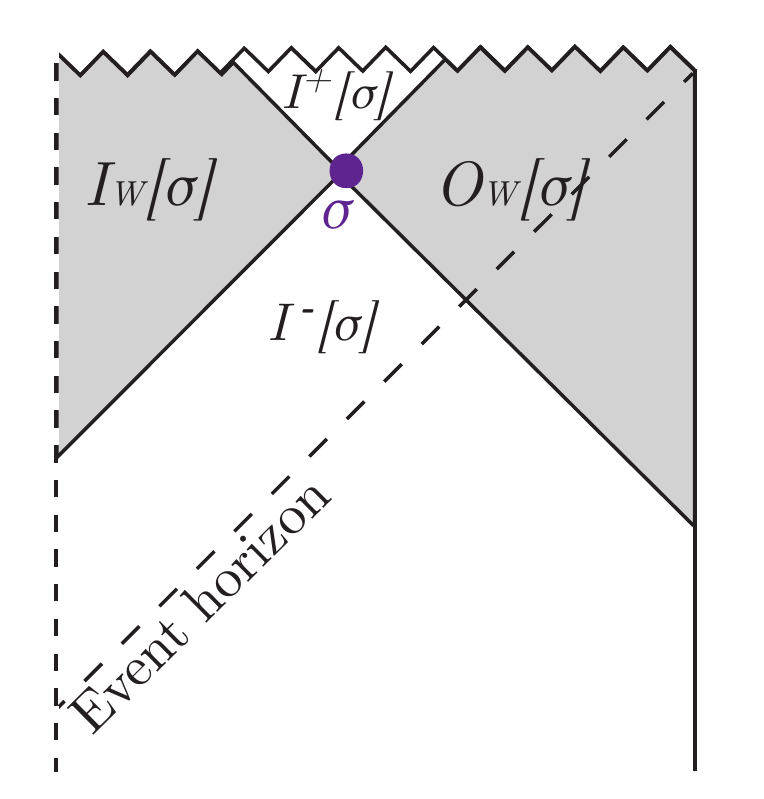}
\caption{\small A conformal diagram of an asymptotically AdS black hole formed from collapse. The figure shows a compact, spacelike, codimension-2 surface $\sigma$ (purple dot), and the decomposition of the spacetime into the future of $\sigma$, $I^{+}[\sigma]$, the past of $\sigma$, $I^{-}[\sigma]$, the exterior of $\sigma$,  $O_{W}[\sigma]$, and the interior of $\sigma$, $I_{W}[\sigma]$.}
\end{center}
\end{figure}

To be a little more precise, any compact surface $\sigma$ that splits a time slice into two pieces induces a natural division of the spacetime into four components: the past of $\sigma$, denoted $I^{-}[\sigma]$, the future of $\sigma$, denoted $I^{+}[\sigma]$, inner wedge of $\sigma$, $I_{W}[\sigma]$ and outer wedge $O_{W}[\sigma]$~\cite{BouEng15b}.  This is illustrated in Fig.~\ref{fig:innerouterdecomp} for a surface inside a black hole.

We will now define the outer entropy as the maximum of the boundary von Neumann entropy given ignorance of the interior:
\be
S^{(\mathrm{outer})}[\sigma] \equiv \max\limits_{\rho\in C} (S_{vN}),
\ee
where the $C$ is the set of all density matrices in the CFT whose classical bulk dual exists and contains the fixed region $O_{W}[\sigma]$. The classical ``microstates'' of $S^{(\mathrm{outer})}[\sigma]$ are all possible spacetime regions that are allowed in $I_{W}[\sigma]$ given that $O_{W}[\sigma]$ is fixed.



But which surface $\sigma$ should we use?  At this point, we need to define more carefully what we mean by the interior of a black hole.  The event horizon of a black hole is defined as the boundary of the region which is inaccesible to future infinity.  Hawking proved that the area of the event horizon is increasing with time \cite{Haw71}.  But the event horizon is teleological, in the sense that its location can depend on what is going to happen in the future.  

There have been a number of proposed definitions of a more local version of a black hole horizon (and consequently, a black hole interior) in classical gravity~\cite{HawEll, Hay93, AshKri02, BouEng15a}.  These definitions all exploit the concept of a marginally trapped surfaces, for which the area of outgoing lightrays is stationary to first order.  Classically, marginally trapped surfaces always lie inside of event horizons.  The local horizons in~\cite{HawEll, Hay93, AshKri02, BouEng15a} are all defined so that they are always foliated by marginally trapped surfaces, satisfying certain additional inequalties.  These definitions are nonunique (on the same black hole spacetime, one can usually find infinitely many surfaces satisfying the criteria), but they do obey laws of thermodynamics similar to the event horizon (see~\cite{HaywardBook} for a review); in particular, these local horizons obey various area-increase theorems \cite{Hay93, AshKri02, BouEng15a}.

In previous work, we showed that the outer entropy of a slice of the event horizon is \emph{not}, in general, given by its area \cite{EngWal17a}; in fact in some situations the outer entropy vanishes, while the area of the event horizon does not.  Although the area of the event horizon is generically greater than the HRT surface \cite{Wal12}, it remains unclear what coarse graining procedure, if any, corresponds to its area \cite{HubRan12,KelWal13}.

More happily, in \cite{EngWal17b} we showed that for an apparent horizon (a codimension-two outermost marginally trapped surface on a time slice) the outer entropy \emph{is} proportional to its area.  This allowed us to explain the area increase theorem for certain spacelike or null holographic screens.  Besides providing a holographic interpretation for the area of non-minimal extremal surfaces, this shows that there is a natural notion of coarse graining associated with the area of the apparent horizon.  We also proposed a boundary dual to the outer entropy of the apparent horizon, called the simple entropy.  This article will give a more detailed and formal version of these arguments.

We extend \cite{EngWal17b} by generalizing the notion of an apparent horizon to a ``minimar surface'' $\mu$, satisfying weaker conditions than an apparent horizon.  In addition to being a compact marginally trapped surface, a minimar surface must satisfy certain minimality inequalities given in Section \ref{sec:Prel}.  The main result of this article is that for a minimar surface, the outer entropy of $\mu$ equals its Bekenstein-Hawking entropy, i.e. the area over four in Planck units:
\be\label{eq:coarse}
S^{(\mathrm{outer})}[\mu] = \frac{\mathrm{Area}[\mu]}{4G\hbar},
\ee
where $S_{vN}$ will be calculated using the HRT formula \eqref{HRTintro}.  

This equality automatically implies a second law for certain kinds of local horizons.  Suppose we have a spacelike (or null) local horizon foliated by minimar surfaces.  We will show from Eq. \eqref{eq:coarse} that $S^{(\mathrm{outer})}[\mu]$ is monotonically increasing as we move spatially outwards, since we are maximizing $S_{vN}$ subject to fewer constraints.  This gives a statistical explanation for the area increase theorem obeyed by such local horizons.

We now briefly summarize our derivation of \eqref{eq:coarse} in the main text.  We must show that, among all bulk states whose classical gravity dual contains $O_{W}[\mu]$, the maximum possibe area of the HRT surface $X$ is equal to the area of $\mu$.  This is done in two steps:
\begin{enumerate}
	\item We show that in any spacetime, the area of the HRT surface is bounded from above by the area of any minimar surface $\mu$.  Thus, even if we vary the interior of $\mu$, the von Neumann entropy remains bounded by the area of $\mu$:
\be\label{eq:ineq}
S[\rho'] \leq \frac{\mathrm{Area}[\mu]}{4G\hbar},
\ee
for any state $\rho'$ with a bulk dual whose outer wedge $O_{W}[\mu]$ agrees with $\rho$.
	\item We explicitly construct an interior for $\mu$ in which this bound is saturated; we do this by patching an interior of $\mu$ to $O_{W}[\mu]$ such that the resulting spacetime has an HRT surface whose intrinsic geometry (and hence area) is the same as that of $\mu$. Because $S^{(outer)}$ is defined as the maximum of the von Neumann entropy, the fact that~\eqref{eq:ineq} is saturated immediately implies that
\be \label{eq:sat}
S^{\mathrm{(outer)}}[\mu]=  \frac{\mathrm{Area}[\mu]}{4G\hbar}.
\ee
\end{enumerate}
Point (1) is a simple consequence of the focusing theorem and the maximin formulation of covariant holographic  entanglement entropy~\cite{Wal12}. Point (2) is more involved; to execute it, we will make use of the initial data problem on characteristic surfaces (i.e. lightfronts) in General Relativity. This will require us to develop junction conditions for gluing data across a codimension-two surface. As a consequence, we also obtain a general procedure for matching two initial data sets in General Relativity at a codimension-two boundary.\footnote{Some techniques in numerical relativity, such as the ``turducken black hole''~\cite{Turducken, Stuffed} or the characteristic-Cauchy matching of~\cite{GomWin97} do use initial data matching across a surface (rather than across a buffer region); but in the former case, the matched initial data is taken to be arbitrary, with allowed violations of the Einstein constraint equations; in the latter case, the matching is adapted specifically to the apparent horizon's exterior (and to finding the apparent horizon). While some case-by-case examples of initial data matching exist (see~\cite{Bona2009} for an example), we are not aware of existing algorithmic junction conditions besides the specialized ones of~\cite{GomWin97}.}

(As a special degenerate case of this construction, we can take our minimar surface to be an extremal surface $X$ which is \emph{not} the one of minimal area (HRT).  This will satisfy our minimar conditions as long as there are no extremal surfaces with lesser area closer to the boundary.  In this case $X = \mu$ and we construct a new spacetime in which $X$ \emph{is} the HRT surface.  This spacetime is simply $O_{W}[X]$ glued to its CPT conjugate.  This provides an interpretation for the area of a class of non-minimal extremal surfaces as the von Neumann entropy in a coarse-grained state, or equivalently the outer entropy in the original state.\footnote{We will also extend this construction to the case of nonminimal extremal surfaces anchored to the boundary, in order to define a coarse-grained entropy for subregions of the CFT.})

So far, our coarse graining has been defined almost entirely on the bulk side.  The only part of the construction which is ``holographic'' was the interpretation of the HRT surface as the fine-grained entropy of the modified spacetime (and hence, as the coarse-grained entropy of the original spacetime).

However, under certain assumptions, we can also provide a boundary dual, the \textit{simple entropy}.  The term ``simple'' denotes operators or sources whose corresponding bulk excitations propagate locally into the bulk\footnote{This is in contrast with complicated nonlocal CFT operators that could modify fields deep in the bulk. For this reason we call these operators ``simple''.}.  (In the classical regime, we can restrict attention to cases where these sources and operators are integrals of local operators, i.e. one-point functions.)  We then define the simple entropy as the maximum of the von Neumann entropy subject to fixing the expectation values of all simple operators $\mathcal{O}$ after some initial time $t_i$, where we are also allowed to turn on arbitrary simple sources $J$ after $t_i$:
\be
S^{\mathrm{(simple)}}[t_i] \equiv \max\limits_{\rho',\,J(t > t_i)} \left[ S[\rho'] \, : \, \langle \mathcal{O}(t > t_{i}) \rangle \mathrm{ \ fixed} \right].
\ee
This simple entropy automatically obeys a second law when the slice $t_i$ is pushed to the future.

We can associate a particular minimar surface to a time slice $t_i$ by following in lightrays from $t_i$ until they reach a marginally trapped surface.  For a black hole near equilibrium, $O_{W}[\mu]$ can be identified with the exterior of the event horizon up to perturbative corrections due to matter falling across the horizon.  As we shall show in Section~\ref{sec:simple}, we can remove this matter by turning on some ``simple'' operators in the bulk, which allows us to measure all of the information in the outer wedge $O_{W}[\mu]$ from the one-point functions on the boundary after time $t_i$.  Since it is not possible to measure the information behind $\mu$ by turning on simple sources, this proves that $S^{\mathrm{(simple)}}[t_i] = S^{\mathrm{outer)}}[\mu]$ at least to all orders in perturbation theory.

This paper is structured as follows: Section~\ref{sec:Prel} introduces assumptions and conventions, reviews some of the relevant geometric constructions, and defines minimar surfaces. In Section \ref{sec:Junctions}, we review the Israel junction conditions and then derive matching conditions for a codimension-two surface. Section~\ref{sec:outer} defines the outer entropy. Section~\ref{sec:main} is the main bulk construction containing the proof that the outer entropy is proportional to area of minimar surfaces. Section~\ref{sec:others} discusses the outer entropy of extremal surfaces and non-minimar surfaces. In Section~\ref{sec:simple}, we define the simple entropy and argue that it is equal to the outer entropy of a minimar surface. Finally, Section~\ref{sec:secondlaw} gives an explanation of the second law for holographic screens foliated by minimar surfaces.  We also motivate a new perspective on how to think about coarse-graining and the second law in ordinary (non-holographic) field theories. Finally, Section~\ref{sec:prospects} discusses the prospects for extending our work beyond classical AdS/CFT.

\section{Preliminaries}\label{sec:Prel}
This section establishes terminology, definitions, and assumptions that will be used throughout the paper.

\subsection{Assumptions, Conventions, and Definitions} \label{sec:Defs}

We will assume the AdS/CFT correspondence, and we work in the large-$N$, large-$\lambda$ limit, in which the bulk $M$ is well-approximated by classical gravity and the RT~\cite{RyuTak06} and HRT~\cite{HubRan07} proposals are valid. We will further assume the Null Convergence Condition (NCC): the requirement that
\be\label{eq:NCC}
R_{ab}k^{a}k^{b}\geq 0,
\ee
where $R_{ab}$ is the spacetime Ricci tensor, for every null vector field $k^{a}$ on $M$. For a spacetime satisfying the Einstein equation, this is equivalent to the Null Energy Condition, which requires positivity of null energy:
\be \label{NEC}
T_{ab}k^{a}k^{b}\geq 0,
\ee
where $T_{ab}$ is the stress-energy tensor in $M$ and as before, $k^{a}$ is any null vector field. \\

We will use the following terminology:
\begin{itemize}
	\item A spacetime $(M,g)$ is a $D$-dimensional Lorentzian manifold, whose metric $g$ is continuous everywhere and smooth almost everywhere.\footnote{i.e. except on a measure zero subset.  This will be important to allow the gluing constructions that are an essential part of this paper.}  For shorthand, we will often refer to $(M,g)$ as just $M$. 
	\item A \textit{surface} will refer to a connected codimension-2 spacelike (embedded) submanifold of $M$ which is compact in the topological interior Int$[M]$. 
	\item A \textit{hypersurface} will refer to a connected codimension-1 (embedded) submanifold of $M$ which is smooth almost everywhere.  A hypersurface will be \textit{splitting} if it divides $M$ into two disjoint components. 
	\item Two surfaces $s_1,s_2$ are \emph{homologous} if there exists a hypersurface $H$ such that $\partial H = s_1 \cup s_2$.
	\item A hypersurface $N$ is \textit{achronal} if no two points on $N$ are timelike separated.
	\item An achronal hypersurface $N$ is \textit{null} if there exists a null vector field $k^{a}$ which is tangent to $N$ at every point where $N$ is smooth. One way of obtaining a null hypersurface is by firing geodesics in a null direction $k^{a}$ from a surface and allowing the geodesics to leave the hypersurface after intersections and caustics. We will call such hypersurfaces null congruences.

	

	\item The \textit{causal future} (past) of $p$, denoted $J^{+}(p)$ ($J^{-}(p)$), is the union of all past- (future) directed causal curves fired from $p$. The \textit{chronological future} (past) of $p$, denoted $I^{+}(p)$ ($I^{-}(p)$) is the union of all past- (future-) directed timelike curves fired from $p$. We can similarly talk about the past or future of a set $S$: $I^{\pm}[S] = \cup_{p\in S} I^{\pm}(p)$.
	\item $M$ is said to be \textit{globally hyperbolic} if there are no closed causal curves in $M$ and for every pair $p$, $q$ in $M$, the intersection $J^{-}(p)\cap J^{+}(q)$ is compact. Note that by this definition, an asymptotically AdS spacetime $M$ fails to be globally hyperbolic. This is easily circumvented by applying this definition to the conformal compactification of $M$~\cite{Ger70} on its asymptotically AdS boundaries.  Thus, in this paper we will refer to such spacetimes as globally hyperbolic. 
	\item The \textit{domain of dependence} of an achronal hypersurface $\Sigma$, denoted $D[\Sigma]$, is the smallest region satisfying the criterion that every timelike curve that enters $D[\Sigma]$ must intersect $\Sigma$.  We will always take $D[\Sigma]$ to be an open set.
	\item A \textit{Cauchy slice} $\Sigma$ of a globally hyperbolic spacetime $M$ is an achronal hypersurface whose domain of dependence is $M$: $D[\Sigma]=M$.  One can also define a Cauchy slice of a globally hyperbolic region $R \subset M$.
	\item A surface $\sigma$ is said to be \textit{Cauchy-splitting} if it divides a Cauchy slice $\Sigma$ into two disjoint components, which we shall call $\mathrm{In}_{\Sigma}[\sigma]$ and $\mathrm{Out}_{\Sigma}[\sigma]$. A Cauchy-splitting surface induces a natural division of the spacetime $M$ into four regions: $I^{+}[\sigma]$, $I^{-}[\sigma]$, $D[\mathrm{In}_{\Sigma}[\sigma]]$ and $D[\mathrm{Out}_{\Sigma}[\sigma]]$~\cite{BouEng15b}. In the introduction, we discussed the \textit{outer wedge} of $\sigma$: $O_{W}[\sigma]\equiv D[\mathrm{Out}_{\Sigma}[\sigma]]$ and the \textit{inner wedge} of $\sigma$:  $I_{W}[\sigma]\equiv D[\mathrm{In}_{\Sigma}[\sigma]]$. Henceforth, we will take all surfaces to be Cauchy-splitting. 
\end{itemize}

\subsection{Geometry of Null Hypersurfaces}\label{sec:NullGeom}
Here we review some properties and definitions in the geometry of null hypersurfaces.

Let $N_{k}$ be a null hypersurface with generating vector field $k^{a}$ in a globally hyperbolic spacetime $(M,g)$. By definition, $k^{a}k_{a}=0$. Let $\ell^{a}$ be a null vector field satisfying $\ell^{a}k_{a}=-1$. The vector field $\ell^{a}$, often called the ``rigging'' vector field, captures a notion of transversality to $N$. Note that in general $\ell^{a}$ is not unique. 

The induced metric on $N_{k}$ is degenerate; the induced metric on spacelike slices of $N_{k}$ orthogonal to $\ell^{a}$ is given by:
\be \label{nullmet}
h_{ab} = g_{ab} +2\ell_{(a}k_{b)}.
\ee
This metric allows us to define the null and transverse extrinsic curvatures of $N_{k}$, respectively:
\begin{align}
& B_{ab}\,_{(k)} = h_{a}^{c}h_{b}^{d} \nabla_{a}k_{b}\\
& B_{ab}\,_{(\ell)}=h_{a}^{c}h_{b}^{d} \nabla_{a}\ell_{b}.
\end{align}

The null extrinsic curvature $B_{ab}\,_{(k)} $ can be decomposed into its trace and traceless parts:
\be \label{eq:Bdecomp}
B_{ab}\,_{(k)}  = \frac{1}{D-2}\theta_{(k)} h_{ab} + \varsigma_{ab}\,_{(k)},
\ee
where $ \varsigma_{ab}\,_{(k)} = B_{ab}\,_{(k)}  - \frac{1}{D-2} \theta_{(k)} h_{ab}$ is a rank-2 tensor that measures the shearing of the congruence with evolution along $k^{a}$; the expansion $\theta_{(k)}=\mathrm{Tr}(B_{ab}\,_{(k)} )$ is a scalar that measures the rate of change the cross-sectional area of $N_{k}$ with evolution along an affinely-parametrized $k^{a}$:
\be
\theta_{(k)}=h^{ab}B_{ab}\,_{(k)} = \frac{1}{2} h^{ab}{\cal L}_{k}h_{ab} = \frac{1}{\sqrt{h}} {\cal L}_{k}\sqrt{h} = \frac{1}{\delta A}\frac{d \delta A}{d\lambda},
\ee
where $\lambda$ is a parameter along the $k^{a}$ geodesics generating $N_{k}$, $\delta A$ is the infinitesimal area element of cross-sections of $N_{k}$, and ${\cal L}_{k}$ is the Lie derivative in the $k^{a}$ direction. The shear and expansion are related via the Raychaudhuri equation\footnote{This equation assumes, as stated in section \ref{sec:Defs}, that $k^a$ is orthogonal to the hypersurface, i.e. that the vorticity $\omega_{ab}$ vanishes.}:
\be
\nabla_{k}\theta_{(k)} =-\kappa_{(k)} \theta_{(k)} -\frac{1}{D-2}(\theta_{(k)})^{2} - \varsigma^{ab}\,_{(k)} \varsigma_{ab}\,_{(k)}-R_{ab}k^{a}k^{b}.
\ee
Here $R_{ab}$ is the \textit{spacetime} Ricci tensor and $\kappa_{k}$ is the inaffinity of the $k^{a}$ congruence: it measures the failure of the $k^{a}$ geodesics to be affinely parametrized: 
\be
k^{a}\nabla_{a}k^{b} = \kappa_{(k)}k^{b}.
\ee
For null geodesic congruences that are affinely parametrized, $\kappa_{(k)}=0$. In such cases, the NCC~\eqref{eq:NCC} is sufficient by the Raychaudhuri equation to guarantee that gravitational curvature can only cause $\theta_{(k)}$ to decrease. The physical interpretation is that gravity satisfying the NCC can only cause light rays to focus, as $A'(\lambda)$ can only decrease. In a spacetime satisfying the Einstein equation, the NEC guarantees that once light rays begin to focus, they must continue to do so: the derivative $k^{a} \nabla_{a}\theta_{(k)}$ is monotonically nonincreasing. 

One final player remains missing: the extrinsic twist potential (a.k.a. the normal fundamental form), \emph{twist} for short. It is a 1-form defined using both $\ell^{a}$ and $k^{a}$: 
\be \chi_{a}\,_{(k)} =\frac{1}{2}h^{c}_{a} \ell^{d} \nabla_{c}k_{d}.
\ee 
Intuitively the twist measures the spacetime dragging of a rotating mass; it is simple to see this in the Lense-Thirring effect in the weak field limit (see e.g.~\cite{Hay06} for a derivation). Note that the twist is antisymmetric under exchange of $\ell$ and $k$: $\chi^{a}\,_{(k)} =-\chi^{a}\,_{(\ell)}$, so it can also be written $\chi_{a} = (\chi_{a}\,_{(k)} -\chi_{a}\,_{(\ell)})/2$.

\subsection{Marginal, Extremal, and Minimar Surfaces}\label{sec:MTS}
\begin{figure}
\centering
\includegraphics[width=0.9\textwidth]{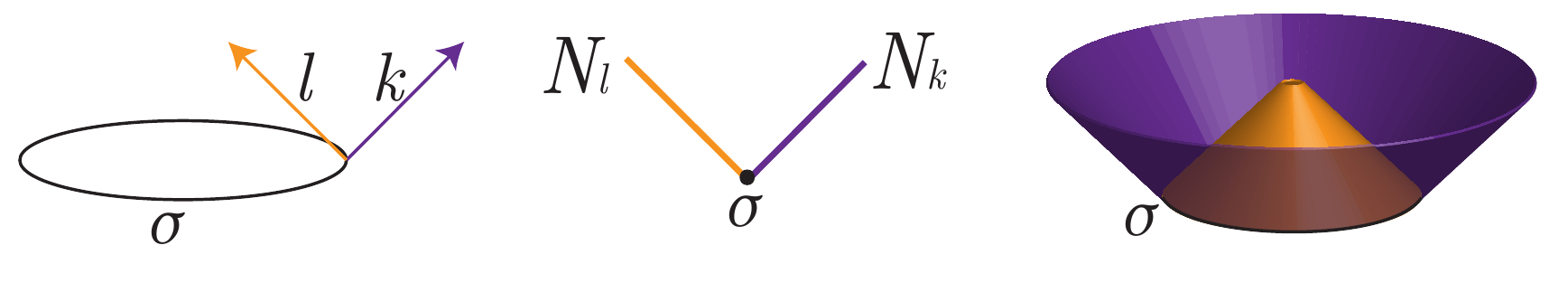}
\caption{A cartoon showing the different ways of denoting the orthogonal null vector fields and hypersurfaces generated from a surface $\sigma$. The left panel figure shows the vectors $\ell^{a}$ and $k^{a}$ at a point on $\sigma$ in $D=3$ dimensions. In the center panel, the null congruences $N_{\ell}$ and $N_{k}$ are shown with $(D-2)$ spacetime dimensions suppressed. The final panel figure shows $N_{\ell}$ (orange) and $N_{k}$ (purple) in $D=3$.}\label{fig:NullDec}
\end{figure}

A surface $\sigma$ has by definition two linearly independent null normals $\ell^{a}$ and $k^{a}$. Along each of these null vector fields, we may fire congruences of null geodesics $N_{\ell}$ and $N_{k}$. These are illustrated in Fig.~\ref{fig:NullDec}.

\begin{figure}
\begin{center}
  \begin{tabular}{  l | c | r }

    \textbf{Surface type} & $\boldsymbol{\theta_{(\ell)}}$ & $\boldsymbol{ \theta_{(k)}}$ \\ 
    \hline \hline
    Untrapped & $-$ & $+$ \\ 
    
    Trapped & $-$ & $-$ \\
    
Marginally Trapped & $-$ & $0$ \\
    
    Anti-Trapped & $+$ & $+$ \\
    
Marginally Anti-Trapped & $+$ & $0$ \\
    
    Extremal & $0$ & $0$
  \end{tabular}
\end{center}
\caption{A table summarizing the classification of surfaces by the expansion of null congruences fired from them. Our conventions are such that whenever one expansion vanishes, we take $k^{a}$ to be the corresponding generating null vector, and for untrapped surfaces $\theta_{(k)}>0>\theta_{(\ell)}$.}
\label{fig:table}
\end{figure}
\begin{figure}
\centering
\includegraphics[width=0.4\textwidth]{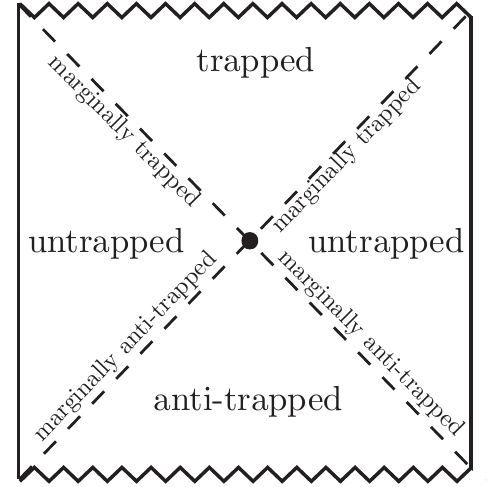}
\caption{A conformal diagram of maximally-extended Schwarzschild-AdS, which contains spherically-symmetric surfaces of all types under the classification of Table~\ref{fig:table}. There are trapped surfaces in the black hole region, anti-trapped surfaces in the white hole region, and untrapped surfaces in each asymptotic region. As with all stationary black holes, the future event horizons are foliated by marginally trapped surfaces; the past event horizons are foliated by marginally anti-trapped surfaces. The bifurcation surface (black dot) is extremal.}
\label{fig:Schw}
\end{figure}

It is convenient to classify surfaces based on the expansions $\theta_{(\ell)}$ and $\theta_{(k)}$ at $\sigma$. When $\theta_{(\ell)}$ and $\theta_{(k)}$ are both positive or both negative, under the right assumptions (including the NCC), we are guaranteed that the spacetime is geodesically incomplete~\cite{Pen65, Haw65,  Haw66, Pen69}. When both expansions are negative, this corresponds to a crunching geometry, where null geodesics are trapped; when both expansions are positive, this corresponds to an expanding geometry, where null geodesics are ``anti-trapped''. ``Untrapped'' surfaces have positive $\theta_{(k)}$ and negative $\theta_{(\ell)}$ or vice versa. The natural boundary between untrapped and trapped or anti-trapped regions are ``marginal'' surfaces with one expansion identically zero on the whole surface. The terminology is summarized in Table~\ref{fig:table}. Fig.~\ref{fig:Schw} illustrates the different types of surfaces in the  Schwarzschild black hole spacetime. Note that we will largely confine the discussion to marginally trapped surfaces (MTSs), with the understanding that the same statements apply in the time reverse to marginally anti-trapped surfaces, which are more useful in  cosmology.

We now launch into a discussion of the most oft-used type of surface in holography:

\paragraph{Extremal Surface:} A surface $X$ is \textit{extremal} if the expansions of the two null orthogonal congruences fired from it both vanish:
\begin{align}\label{eq:expansion}
&\theta_{(\ell)}= 0\\
&\theta_{(k)}= 0. \label{eq:expansion2}
\end{align}
Since any vector orthogonal to $X$ can be written as a linear combination of $\ell^{a}$ and $k^{a}$, it immediately follows that the area of $X$ is stationary to first order perturbations in any direction. The \textit{HRT surface} of a connected component of the asymptotic boundary $B$ is  the minimal area surface homologous to $B$ satisfying Eq.~\eqref{eq:expansion} \& \eqref{eq:expansion2}.

The HRT prescription~\cite{RyuTak06, HubRan07, LewMal13, DonLew16} for computing the von Neumann entropy of an entire connected component $B$ of the CFT at one time is the following formula:
\begin{equation}\label{eq:HRT}
S_{vN}= -\mathrm{tr} (\rho_{B} \ln \rho_{B})=\frac{\text{Area}[X]}{4 G \hbar},
\end{equation}
where $\rho_{B}$ is the density matrix of $B$, and $X$ is the minimal area extremal surface homologous to $B$.\footnote{The HRT prescription can also be used to calculate the von Neumann entropy of subregions of the boundary that do not constitute a complete connected component, but except in Sec.\ref{sec:extremal} our results will be shown in the case where $R = B$.  However, we believe that much of what we say could be extended to the case of a general region $R$.}  Note that in the case of a one-sided black hole (e.g. a black hole formed from collapse), $X$ is given by the empty set so $S_{vN} = 0$ at classical order.

An equivalent formulation of the HRT surface of which we will make frequent use is the maximin construction~\cite{Wal12}. In the maximin construction, one first identifies of the minimal area surface homologous to  $B$ on a given Cauchy slice $\Sigma$; we shall denote this surface by min$(B,\Sigma)$. One then chooses $\Sigma$ so as to maximize the area of the minimal area surface over all possible Cauchy slices.  Using the NCC together with some global assumptions,\footnote{This excludes spacetimes featuring e.g. an inflating de Sitter asymptotic region behind the horizon (see~\cite{FisMar14}), where maximin/HRT surfaces do not necessarily exist in the (real, nonconformally compactified) spacetime geometry, and the holographic interpretation is unclear.} one can show that this is equivalent to the HRT surface~\cite{Wal12}.  The following is a  very useful consequence of the maximin formalism:

\paragraph{Lemma:}\cite{Wal12} An HRT surface $X$ is the minimal area surface homologous to $B$ on some Cauchy slice containing $X$.\\

The region between $X$ and $B$ is commonly referred to as the entanglement wedge:

\paragraph{Entanglement Wedge:} The entanglement wedge $E_{W}[B]$ of $B$, referred to also as the exterior of the HRT surface $X$, is defined as the domain of dependence of any hypersurface $\text{Out}_{\Sigma}[X]$ connecting $X$ to $B$~\cite{CzeKar12, Wal12, HeaHub14}:
\be
E_{W}[B]=D[\text{Out}_{\Sigma}[\sigma]].
\ee
It is now understood that $E_{W}[B]$ is the region dual to the CFT density matrix $\rho_{B}$, so that field data in $E_{W}[B]$ can be fully reconstructed from operators in $B$, and it commutes with operators on the complementary boundary $\widetilde{B}$~\cite{JafLew15, DonHar16}. \\

More generally, for any surface $\sigma$ homologous to $B$,  we can give a natural generalization of the entanglement wedge to the region bounded between arbitrary surfaces $\sigma$ and $B$, which we shall call the \textit{outer wedge} of $\sigma$:

\paragraph{Outer Wedge:} Let $\sigma$ be a surface homologous to $B$. Let $\Sigma$ be a Cauchy slice containing $\sigma$, with decomposition into disjoint components as given in Sec.~\ref{sec:Defs}.  Then $\Sigma=\text{In}_{\Sigma}[\sigma]\cup \sigma \cup \text{Out}_{\Sigma}[\sigma]$ where $\text{Out}_{\Sigma}[\sigma]$ is any homology slice connecting $\sigma$ to $B$. The outer wedge of $\sigma$, denoted $O_{W}[\sigma]$ is the domain of dependence of $\text{Out}_{\Sigma}[\sigma]$:
\be
O_{W}[\sigma]=D[\text{Out}_{\Sigma}[\sigma]].
\ee

We define the inner wedge of a surface in an analogous way, as the domain of dependence of $\text{In}_{\Sigma}[\sigma]$:

\paragraph{Inner Wedge:} Let $\sigma$ be as above. The inner wedge of $\sigma$ is defined as the domain of dependence of $\text{In}_{\Sigma}[\sigma]$:
\be
I_{W}[\sigma] = D[\text{In}_{\Sigma}[\sigma]].
\ee

\begin{figure}
\centering
\includegraphics[width=0.4\textwidth]{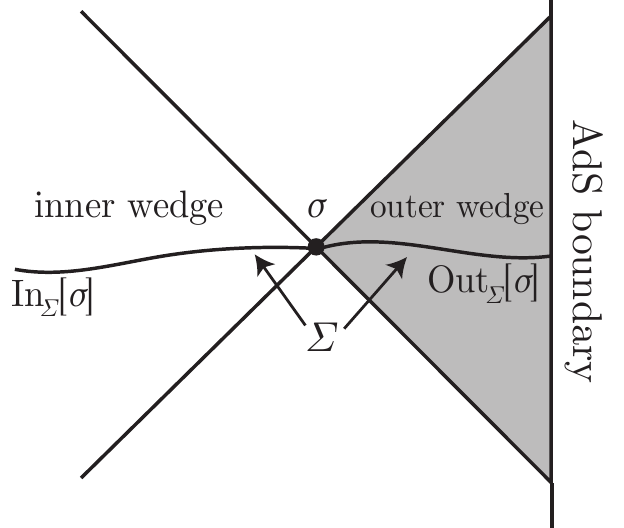}
\caption{Decomposition of the outer and inner wedges of $\sigma$. $\Sigma$ is a Cauchy slice of the full spacetime, and In$_{\Sigma}[\sigma]$, Out$_{\Sigma}[\sigma]$ are the components of $\Sigma$ as split by $\sigma$. }
\label{fig:OutIn}
\end{figure}

The outer and inner wedges of a surface $\sigma$ are illustrated in Fig.~\ref{fig:OutIn}. Note that the union of the outer and inner wedges with $\sigma$ necessarily contains a complete Cauchy slice of the spacetime: specifying data in the two wedges is sufficient to fix the entire spacetime, and the data can be independently specified so long as one solves the constraint equations across $\sigma$.  Note that spacetime points that are timelike to $\sigma$ do not lie in either wedge; these are the points that are causally related to both wedges.

We now consider a natural generalization of extremal surfaces: namely marginal surfaces. Rather than requiring stationarity of the area in both orthogonal null directions as is the case when the surface is extremal (cf. Eq.~\eqref{eq:expansion}\&\eqref{eq:expansion2}), marginal surfaces are only required to be stationary in one null direction.

\paragraph{Marginal Surface:} A surface $\mu$ is \textit{marginal} if the expansions of the two null orthogonal congruences fired from $\mu$ satisfy:
\begin{align}
&\theta_{(\ell)}\leq 0 \mathrm{ \ \ or \ \ } \theta_{(\ell)}\geq 0 \label{def:marNeg}\\ 
&\theta_{(k)}= 0
\end{align}
\noindent where the degenerate case in which $\theta_{(\ell)}=0$ is simply the situation in which $\mu$ is an extremal surface. The first equation requires $\theta_{(\ell)}$ to have the \textit{same} sign on all of $\mu$; i.e. the ``or'' is exclusive. When $\theta_{(\ell)}\leq 0$, $\mu$ is said to be marginally trapped, and when $\theta_{(\ell)}\geq 0$, $\mu$ is said to be marginally anti-trapped.


Guided by intuitions from the the entanglement picture, we define a \textit{minimal} marginal surface, or minimar for short; as in the case of the HRT surface, the area of this surface will turn out to measure the entropy associated with ignorance about its interior.  Whereas the HRT surface measures the fine-grained entropy of $B$ (i.e. the only ignorance is that of anything outside of $B$), this defines a coarse-grained entropy of $B$ (i.e. we are also forgetting some of the information in $B$ itself).

\paragraph{Minimar surface:} A marginal surface $\mu$ will be called a \textit{minimar surface} if it additionally satisfies the following criteria:
\begin{enumerate}
	\item $\mu$ is homologous to $B$, and there exists a Cauchy slice $\Sigma_\text{min}[\mu]$ of $O_W[\mu]$ on which $\mu$ is a minimal area surface homologous to $B$. \label{def:minimarMin}
	\item There exists a choice of normalization for $\ell^a$ such that $\nabla_{k}\theta_{(\ell)}\equiv k^{a}\nabla_{a}\theta_{(\ell)}\le0$ on $\mu$, with equality allowed only if $\theta_{(\ell)} = 0$ everywhere on $\mu$. \label{def:minimarCross}
 \end{enumerate}
		
Condition (1) is a weaker version of the global HRT minimality: we do not require that $\mu$ be the minimal area marginal surface homologous to the boundary (this will in general not be well-defined); instead we require that it be minimal on a partial Cauchy slice.

Condition (2) may appear to be a new additional condition with no parallel in the minimality condition of HRT surfaces. However, we will prove in  Appendix~\ref{sec:HRT} that $\nabla_{k}\theta_{(\ell)}\leq 0$ on HRT surface with equality being highly nongeneric.  When $\mu$ is only marginal, we must impose condition (2) separately. Condition (2) is also known as ``strict spacetime stability'' of a marginal surface~\cite{AndMar05}; it guarantees that small deformations of the surface inwards in a null direction can result in a trapped (anti-trapped) surface, while small deformations outwards in a null direction can result in an untrapped surface. 

It is possible to prove that generic apparent horizons are minimar surfaces, so that our results apply to generic apparent horizons, the case originally investigated in our earlier work. 


%
%
%
%
%
%
%
%
%


\section{Junction Conditions for Initial Data}\label{sec:Junctions}

\begin{figure}[t]
	\centering
	\subfigure[]{\includegraphics[width=0.4\textwidth]{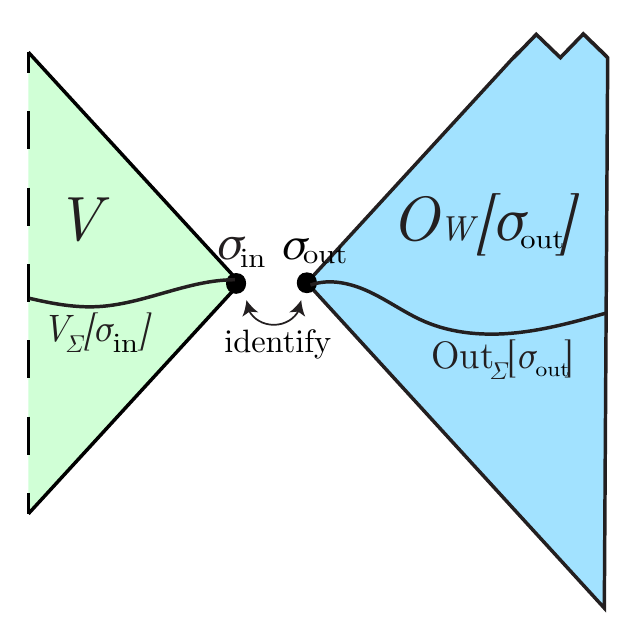}
		\label{subfig:ID}
	}
	\hspace{2cm}
	\subfigure[]{\includegraphics[width=0.4\textwidth]{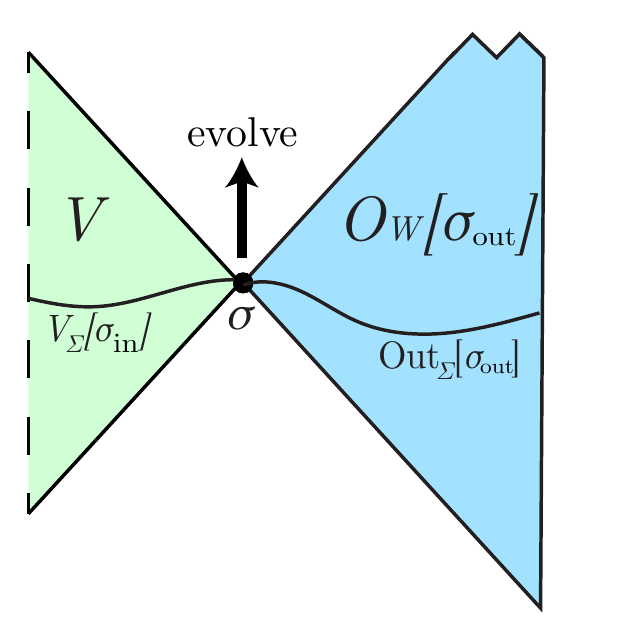}
		\label{subfig:evolve}
	}
	\caption{A cartoon illustrating the construction of gluing conditions for an interior of a codimension-2 surface $\sigma_{\text{out}}$. \ref{subfig:ID} The first step of the construction describes when a spacetime $V$ with Cauchy slice $V_{\Sigma}[\sigma_{\text{in}}]$ whose boundary is $\sigma_{\text{in}}$ can be consistently identified with $O_{W}[\sigma_{\text{out}}]$ across $\sigma_{\text{in}}$; this requires topological consistency across $\sigma_{\text{out}}$ to $\sigma_{\text{in}}$ as well as data sufficiently smooth that $V_{\Sigma}[\sigma_{\text{in}}]\cup \text{Out}_{\Sigma}[\sigma_{\text{out}}]$ satisfies the Einstein constraint equations. \ref{subfig:evolve}: the data on the slice $V_{\Sigma}[\sigma_{\text{in}}]\cup \text{Out}_{\Sigma}[\sigma_{\text{out}}]$ is then evolved via the equations of motion. }
	\label{fig:outline}
\end{figure}

\begin{figure}[t]
	\centering
	\subfigure[]{\includegraphics[width=0.45\textwidth]{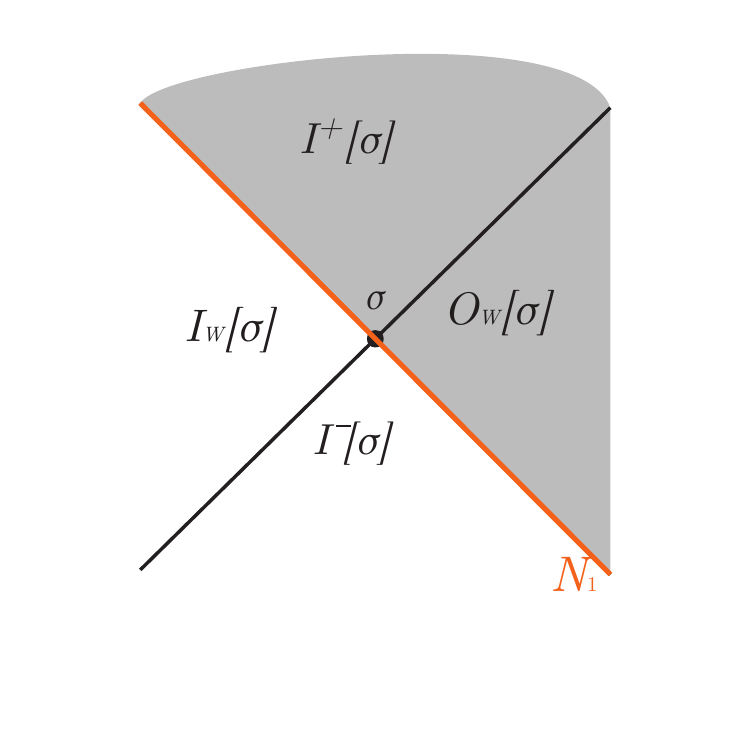}
		\label{subfig:N1}
	}
	\hspace{1cm}
	\subfigure[]{\includegraphics[width=0.45\textwidth]{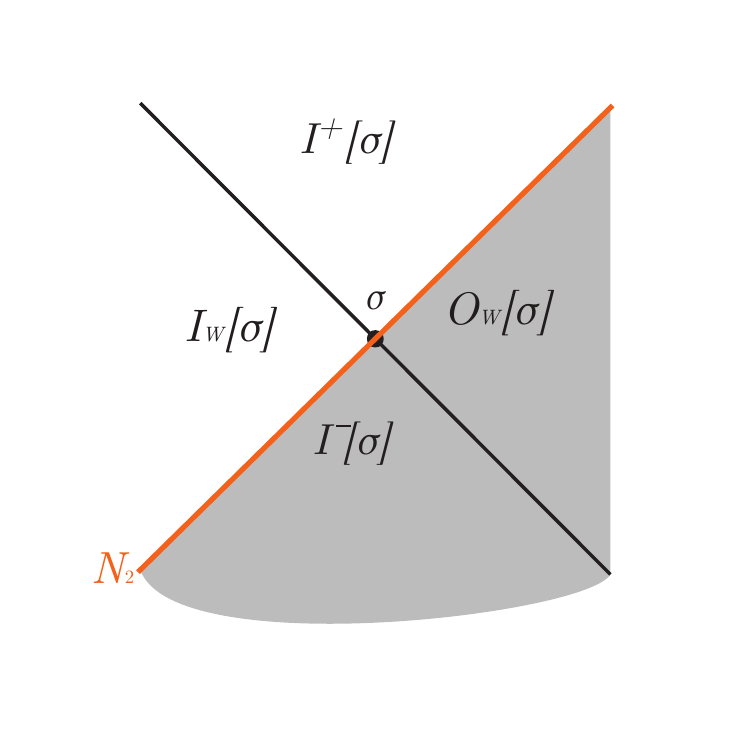}
		\label{subfig:N2}
	}
	\caption{The junction conditions for gluing together two globally hyperbolic spacetime regions $I_{W}[\sigma]$ and $O_{W}[\sigma]$ across a codimension-two hypersurface are derived via a twofold application of junction conditions across a null hypersurface. We postulate the existence of $I^{\pm}[\sigma]$; \ref{subfig:N1} shows the first set of junction conditions, obtained by dividing the spacetime up so that $I^{+}[\sigma]$ and $O_{W}[\sigma]$ are on the same side of the null hypersurface $N_{1}$ (orange); \ref{subfig:N2} shows the second set of junction conditions, obtained by dividing the spacetime so that $O_{W}[\sigma]$ and $I^{+}[\sigma]$ are on opposite sides of the null hypersurface $N_{2}$ (orange). }
	\label{fig:Ns}
\end{figure}

A description of the possible spacetimes that can constitute $I_{W}[\sigma]$ requires a set of conditions that dictate whether a given spacetime region $V$ can be sewed onto $\sigma$ in such a way that the resulting patched spacetime of $V$ and $O_{W}[\sigma]$ is a manifold with a continuous metric and a well-behaved causal structure, which solves the distributionally-well defined Einstein equation with a stress-energy tensor that satisfies the NEC. 

The procedure is twofold: first, the region $V$ is patched onto $O_{W}[\sigma]$, then the initial data on a Cauchy slice of the patching (see Fig.~\ref{fig:outline} for an illustration) must be evolved to give rise to a new spacetime $\widetilde{M}$. Note that because $V$ and $O_{W}[\sigma]$ are by definition spacelike-separated, they must separately satisfy the Einstein equation; the constraints on $V$ must come from the junction at $\sigma$ itself. 


The task at hand is thus a problem of both junction conditions of spacetime regions and initial data engineering. The problem of gluing together two spacetimes satisfying the Einstein equation has been studied extensively for junctions across codimension-one hypersurfaces~\cite{Dar27, ObrSyn52, Lic55, Isr58, Isr66, Rob72, BonVic81, ClaDra87, BarIsr91, MarSen93}, but as far as we are aware, gluing across a codimension-two surface has received relatively little attention: initial data set patching conditions are normally given via an intermediate region rather than over a surface.

To derive the constraints imposed on $I_{W}[\sigma]$ by $O_{W}[\sigma]$, we instead employ a twofold application of the junction conditions across two codimension-one null hypersurfaces, and then invoke the initial data formulation of general relativity. A rough sketch is as follows: let $\widetilde{M}$ be a consistent spacetime containing $O_{W}[\sigma]$. Using the decomposition of $\widetilde{M}$ induced by $\sigma$, we can divide $\widetilde{M}$ into two spacetime regions: $O_{W}[\sigma]\cup J^{+}[\sigma]$ and $I_{W}[\sigma]\cup J^{-}[\sigma]$. This is illustrated in Fig.~\ref{fig:Ns}. As explained above, the two regions are separated by a null hypersurface $N_{1}$; the Barrab\`es-Israel junction conditions~\cite{BarIsr91} give the requisite constraints on $N_{1}$ for $\widetilde{M}$ to be consistent. Repeating the procedure, but this time breaking up $\widetilde{M}$ into $O_{W}[\sigma]\cup J^{-}[\sigma]$ and $O_{W}[\sigma]\cup J^{+}[\sigma]$. This gives conditions that must be satisfied by $N_{2}$. Together, these give junction conditions at the intersection $N_{1}\cap N_{2}=\sigma$. The conditions give a precise constraint on the spacetimes that are allowed to be inside $\sigma$.  Finally, to obtain a full spacetime, we invoke the initial data formulation of general relativity on an achronal slice containing $\sigma$; this guarantees that $\widetilde{M}$ exists\footnote{Technically, the proofs of the initial data problem guarantee only local existence and normally assume a somewhat higher differentiability order than we do. We will discuss these subtleties in Sec.~\ref{sec:MultiJunction}.}.


\subsection{Review: the Barrab\`es-Israel Junction Conditions} \label{sec:Israel}

\begin{figure}
\centering
\includegraphics[width=0.45\textwidth]{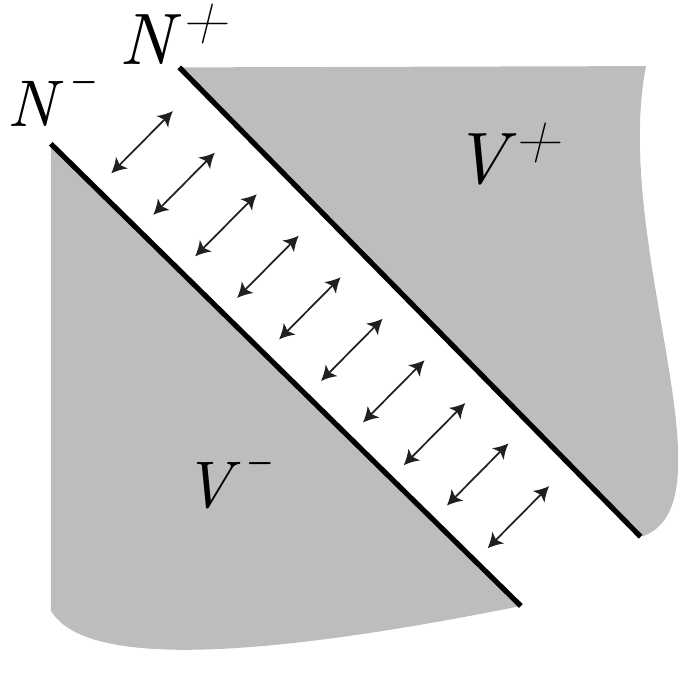}
\caption{The Barrab\`es-Israel junction conditions prescribe when two spacetime regions $V^{+}$, $V^{-}$ with null boundaries $N^{+}$, $N^{-}$ can be sewn together by identifying $N^{+}$ with $N^{-}$.}
\label{fig:sewing}
\end{figure}

Let $(M^{+},g^{+})$, $(M^{-},g^{-})$ be two $C^{3}$ globally hyperbolic spacetimes satisfying the Einstein equation, and let $N^{+}\subset M^{+}$, $N^{-}\subset M^{-}$ be two splitting null hypersurfaces. Let $V^{+}=J^{+}[N^{+}]$ and $V^{-}=J^{-}[V^{-}]$. These are illustrated in Fig.~\ref{fig:sewing}.
Suppose that we wanted to construct a new spacetime by identifying $N^{+}$ and $N^{-}$. What conditions must be imposed across the junction so the new geometry satisfies the Einstein equation? 

The most basic requirement for a patched spacetime of $V^{+}$ and $V^{-}$ across $N^{+}$ and $N^{-}$ is that the resulting set be smooth as a topological manifold: $N^{+}$ and $N^{-}$ must be diffeomorphic so that they can be identified as one (embedded) submanifold $N$ of a joined smooth topological space $M\equiv V^{+} \cup V^{-}$. The second requirement is that this submanifold $N$ have a well-defined intrinsic geometry. This requires $\left . h_{ab}\right |_{N^{+}}=\left . h_{ab}\right |_{N^{-}}$, where $h_{ab}$ is the induced metric on a spatial slice of $N$. As differences in quantities across $N$ will be appearing a lot, we will use the standard convention to denote them:
\be
[F] \equiv  \left . F\right |_{N^{+}} - \left. F\right|_{N^{-}},
\ee
where $F$ is any spacetime field. In this convention, $[h_{ab}]=0$. Thus $N^{-}$ and $N^{+}$ must be isometric. This is the first junction condition:\\

\noindent \textit{First Junction Condition:} The null hypersurfaces $N^{+}\subset M^{+}$ and $N^{-}\subset M^{-}$ are isometric (with respect to their induced metrics from $g^{+}$ and $g^{-}$, respectively). \\

A theorem by Clarke and Dray~\cite{ClaDra87} then guarantees that the joined topological space $M$ is a smooth manifold with metric $g$ which is continuous on all of $M$ and $C^{2}$ everywhere except possibly on $N$ itself.
 
Recall that the end goal is a spacetime that satisfies the Einstein equation. To understand the conditions imposed by the Einstein equation, we must study derivatives of the metric across $N$. 

We expect that we can always choose the vectors $\ell^{a}$ and $k^{a}$ to be continuous when $V^{\pm}$ are sufficiently regular and the first junction condition is satisfied: $[\ell^{a}]=[k^{a}]=0$; furthermore, derivatives of $g_{ab}$ along directions tangential to $N$ are continuous as well. In particular:
\be
[g_{ab,c}] k^{c}=0.
\ee
Any junction conditions would therefore have to result from derivatives in the transverse direction $\ell^{a}$, as defined in Sec.~\ref{sec:NullGeom}. The quantity of interest is therefore the change of $g_{ab,c}\ell^{c}$ across $N$. As this is the primary quantity of study, it is worthwhile to give it a name:
\be
\gamma_{ab} \equiv [g_{ab,c}]\ell^{c}. 
\ee

Let us now pause and ask what behavior would be desirable for us to call $(M,g)$ a physical spacetime. At a minimum, the stress-energy tensor sourcing this geometry should be well-defined as a distribution: the worst singularities allowed would be Dirac $\delta$-functions. However, we will be stricter and require all stress-energy tensors to be finite, while still allowing finite discontinuities.  

We now ask the question: what are the contributions of $\gamma_{ab}$ to the Einstein equation?  This requires a straightforward if tedious computation of the discontinuities in the connection coefficients across $N$:
\be
[\Gamma^{a}_{bc}] = k_{c}\gamma^{a}_{b} +k_{b}\gamma^{a}_{c} -k^{a}\gamma_{bc},
\ee
which allows us to compute the discontinuities in the stress-energy tensor via the Einstein equation. The expression is easiest to parse in terms of geometric quantities of the null congruence generated by the $\ell^{a}$ vector field on a spacelike slice $S$ of $N_{k}$. In terms of the expansion $\theta_{(\ell)}$ and twist $\chi_{(\ell) \, a}$ of the transverse null congruence $N_{\ell}$ and the inaffinity $\kappa_{(k)}$ of the null congruence $N_{k}$:\\

\noindent \textit{The Second Junction Condition~\cite{Dar27, ObrSyn52, Lic55, Isr58, Isr66, Rob72, BonVic81, ClaDra87, BarIsr91, MarSen93}:}
\be \label{eq:stressShell}
 T_{ab}(x) ^{\mathrm{shell}}=
  - ([\theta_{(\ell)}] k_{a}k_{b} +[\chi_{(\ell)\,(a}]k_{b)} +[\kappa_{(k)}] h_{ab})
\ee
here $T_{ab}^{\mathrm{shell}}$ is the stress-energy tensor that supports the junction. It is also possible to rewrite $T_{ab}$ in terms of intrinsic coordinates on $N_{k}$; that form is independent of the choice of $\ell^{a}$~\cite{MarSen93}. 

 The reader may notice that not all components of $\gamma_{ab}$ must vanish for the stress tensor to be finite; in particular, the shear $\varsigma$ of either $N_{\ell}$ or $N_{k}$ does not appear in the above equation. This is special to a junction across a null hypersurface; if there is nonvanishing shear across the junction, the spacetime may include an impulsive gravitational wave (which is not sourced by $T_{ab}$)~\cite{Pen72}. 

The stress-energy tensor on $N_{k}$ has a physical interpretation as a surface layer of a shell of null matter. If $[\theta_{(\ell)}]+ [\kappa_{(k)}]\geq 0$, this shell will satisfy the NEC (Eq.~\ref{NEC}). Such shells are used to construct new solutions in General Relativity, including in the context of AdS/CFT (see e.g.~\cite{FreHub05, FisMar14}).

If we now demand that the stress tensor be finite, we obtain the following junction conditions:
\be
[\theta_{(\ell)}] = [\chi_{(\ell)\,a}] =[\kappa_{(k)}]=0
\ee   
The last condition in particular guarantees that $N$ is an affinely parametrizable null geodesic congruence in the full patched spacetime $M$.

\subsection{Multiple Junctions}\label{sec:MultiJunction}
\begin{figure}
\centering
\includegraphics[width=\textwidth]{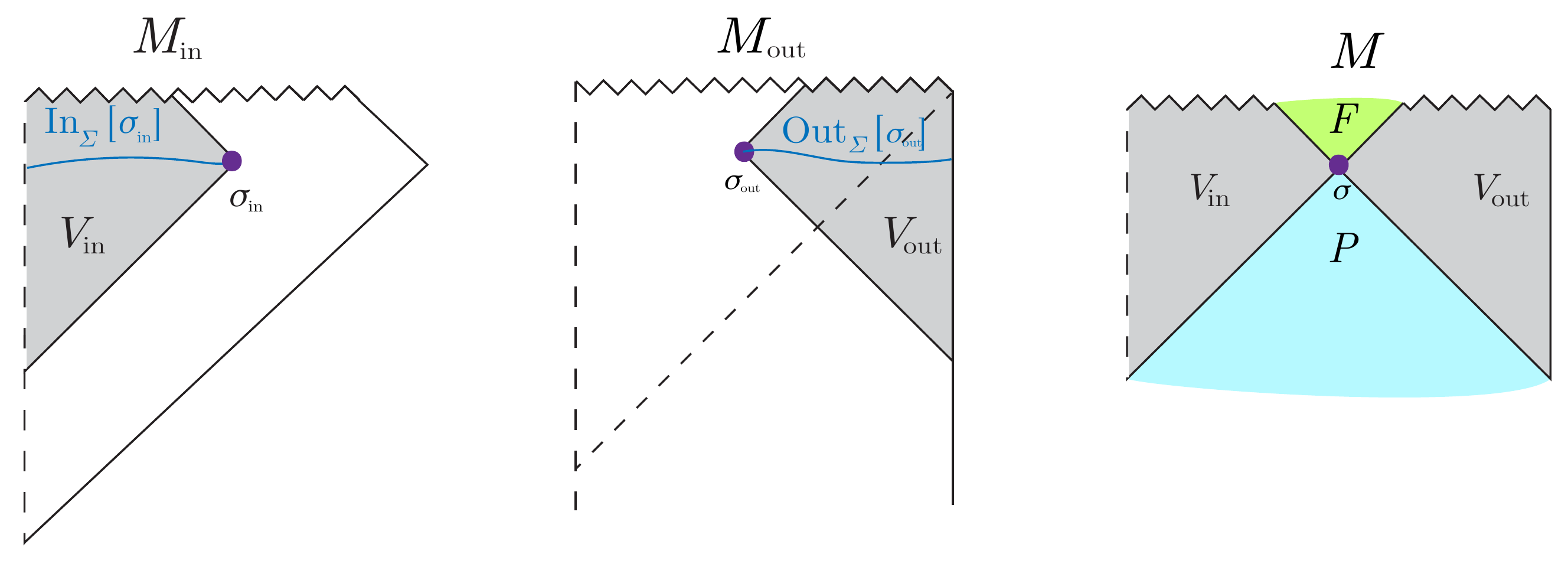}
\caption{The patching construction illustrated in detail. On the left panel, a surface $\sigma_{\text{in}}$ (purple) splits a Cauchy slice $\Sigma_{\text{in}}$ (not shown) in two. The side In$_{\Sigma_{\text{in}}}[\sigma_{\text{in}}]$ is on the interior of $\sigma_{\text{in}}$; the region $V_{\text{in}}$, which is the interior of $\sigma_{\text{in}}$, is obtained by taking the domain of dependence of In$_{\Sigma_{\text{in}}}[\sigma_{\text{in}}]$. The middle panel illustrates the same construction in $M_{\text{out}}$ for $\sigma_{\text{out}}$. The right panel shows the gluing, with $F$ and $P$ the fiduciary spacetime regions that exist only when the junction conditions  are satisfied.}
\label{fig:Multi}
\end{figure}

We can now make use of the Barrab\`es-Israel junction conditions to derive conditions on initial data matching. Instead of taking $V^{+}$ and $V^{-}$ to be spacetime regions in the future and past of a null hypersurface, we take $V_{\mathrm{out}}$ and $V_{\mathrm{in}}$  to be domains of dependence of initial data in $M_{\text{out}}$ and $M_{\text{in}}$. More precisely, let $\Sigma_{\text{out}}, \ \Sigma_{\text{in}}$ be Cauchy slices of $M_{\text{out}},\ M_{\text{in}}$ (which are maximally extended) and let $\sigma_{\text{out}}\subset \Sigma_{\text{out}}$, $\sigma_{\text{in}}\subset \Sigma_{\text{in}}$ be two surfaces, as defined in Sec.~\ref{sec:Defs}. We take $V_{\text{out}}$, $V_{\text{in}}$ to be the domains of dependence of one side of $\Sigma_{\text{out}}$, $\Sigma_{\text{in}}$ each: 
\bea \label{Vpm}
& V_{\mathrm{out}}= D\left [\mathrm{Out}_{\Sigma_{\text{out}}}[\sigma_{\text{out}}]\right]\\
& V_{\text{in}} = D\left [\mathrm{In}_{\Sigma_{\text{in}}}[\sigma_{\text{in}}]\right]
\eea
where In and Out are chosen arbitrarily. This is illustrated in Fig.~\ref{fig:Multi}. Suppose now that we want to patch $V_{\mathrm{out}}$ onto $V_{\text{in}}$ across the \textit{codimension-2} surface $\sigma_{\text{out}}$, $\sigma_{\text{in}}$. To determine the appropriate conditions at this surface, we will make use of the Barrab\`es-Israel junction conditions twice. 

Let us imagine that there is some fiducial spacetime region $F$ such that $F=J^{+}[\sigma_{\text{out}}]$, and some fiducial spacetime $P$ such that $P=J^{-}[\sigma_{\text{in}}]$. What conditions must $ F\cap P$ satisfy so that the entire spacetime $F\cup V_{\mathrm{out}}\cup P \cup V_{\text{in}}$ is consistent and satisfies the Einstein equation?

In order for the topological space $F\cup V_{\mathrm{out}}\cup P \cup V_{\text{in}}$ to be a manifold with a continuous metric, the boundaries of all touching sets must be isometric by the Clarke-Dray theorem. This is equivalent to requiring that $\sigma_{\text{out}}$ and $\sigma_{\text{in}}$ be isometric. This is our first multi-junction condition; we now identify $\sigma_{\text{out}}$ and $\sigma_{\text{in}}$ as a single surface $\sigma$. 

The second condition requires an application of the Barrab\`es-Israel junction conditions twice. First, we consider joining $V_{\mathrm{out}}\cup F$ and $V_{\text{in}}\cup P$ along their mutual boundary. Let us call this hypersurface $N_{k}$, and as above we require that the generator $k^{a}$ be $C^{0}$ across $\sigma$. We pick the rigging vector $\ell^{a}$ so it is normal to $\sigma$ and $C^{0}$. Then Eq.~\ref{eq:stressShell} tells us that in order to have a a regular stress-energy tensor in the null-null directions, we must require that $\theta_{(\ell)}$, $\chi_{a}\,_{(\ell)}$, and $\kappa_{(k)}$ all be continuous across $N_{k}$. Next, we consider joining $V_{\text{out}}\cup P$ and $V_{\text{in}}\cup F$ along the new boundary, which is the null hypersurface generated by the vector $\ell^{a}$, which we take to be $C^{0}$ across $\sigma$ as well. The Barrab\`es-Israel junction condition requires that $\theta_{(k)}$, $\chi_{a}\,_{(k)}$, and $\kappa_{(\ell)}$ all be continuous across $N_{\ell}$. This is illustrated in Fig.~\ref{fig:Ns}. However, unlike in the case of codimension-one hypersurfaces, the condition on the inaffinities is actually vacuous: by an appropriate rescaling, we can always pick the inaffinity to be continuous at $\sigma$.

Because we now consider differences when crossing two null hypersurfaces simultaneously at a codimension-2 surface, the symbol $[F]$ will now denote
the discontinuities of a quantity $F$ across $\sigma$ in crossing from $O_{W}[\sigma]$ to $I_{W}[\sigma]$ in the following way:
\be
[F] \equiv  \left . F\right |_{\sigma_{\text{out}}} -  \left . F \right |_{\sigma_{\text{in}}}.
\ee
Finally, the conditions on $V_{\mathrm{out}}$ and $V_{\text{in}}$ are:

\paragraph{Codimension-Two Junction Conditions:} Let $(V_{\mathrm{out}},g_{\text{out}})$, $(V_{\text{in}}, g_{\text{in}})$ be defined as in~\eqref{Vpm}: $g_{\text{out}}$, $g_{\text{in}}$ are smooth. Then we may glue $V_{\mathrm{out}}$ and $V_{\text{in}}$ to one another with a finite stress-energy tensor under the following conditions:
\begin{enumerate}
	\item The surfaces $\sigma_{\text{out}}$ and $\sigma_{\text{in}}$ are isometric and can thus be identified as a single surface $(\sigma, h)$. 
	\item There exists a choice of $k^{a}$ and $\ell^{a}$ null normals (satisfying Eq.~\eqref{nullmet}) defined on both sides of $\sigma$ such that the following conditions hold: 
\bea 
&  [\theta_{(k)}]=0  \label{contThK}\\
& [\theta_{(\ell)}] =0 \label{contThL}\\
& [\chi_{a\,_{(k)}}] =- [\chi_{a}\,_{(\ell)}]=0 \label{contChi}
\eea
for some $k^{a}$ and $\ell^{a}$ that are $C^{0}$ on $N_{k}$ and $N_{\ell}$ respectively. As before, no continuity condition is imposed on the shear.
 \end{enumerate} 
 \vspace{0.25cm}
 Then the null-null components of the stress tensor are finite, and the Einstein equation is distributionally well-defined. Now, because the data on $\mathrm{Out}_{\Sigma_{\text{out}}}[\sigma_{\text{out}}]$ and on $\mathrm{In}_{\Sigma_{\text{in}}}[\sigma_{\text{in}}]$ is guaranteed to satisfy the constraint equations separately, the conditions Eqs.~\eqref{contThK}- \eqref{contChi} guarantee that the entire slice $\Sigma=\mathrm{Out}_{\Sigma_{\text{out}}}[\sigma_{\text{out}}]\cup \sigma \cup\mathrm{In}_{\Sigma_{\text{in}}}[\sigma_{\text{in}}]$ satisfies the constraint equations with a finite stress-energy tensor. Because there is no contribution to the stress-energy tensor from $\sigma$, the initial data on $\Sigma$ has a stress-energy tensor that satisfies the NEC whenever $(M_{\text{out}},g_{\text{out}})$ and $(M_{\text{in}},g_{\text{in}})$ do. \footnote{As a sanity check, one could test our initial data against the null constraint equations on $N_{k}$. Eqs.~\eqref{contThK}- \eqref{contChi} and continuity of quantities tangent to $N_{k}$ and $N_{\ell}$ together with the constraint equations imply that the stress-energy tensor has at most step-function discontinuities, as desired.}

We should now ensure that the data available is sufficient to prescribe a Cauchy evolution of our data into the ``fiducial'' spacetime regions $F$ and $P$. It is simplest to see that the specified data is sufficient via the characteristic initial data formalism, in which data is specified on a piecewise null hypersurface and evolved forward, as illustrated in Fig.~\ref{fig:outline}. This differs from the standard Cauchy evolution, which requires a smooth spacelike hypersurface. For our purposes here, the characteristic initial data problem of interest is that of two intersecting cones: $N_{k}$ and $N_{\ell}$ intersect on $\sigma$. The requisite geometric data for a characteristic initial data evolution is precisely the data at hand: a conformal metric on $N_{k}$ and $N_{\ell}$, an intrinsic metric $h_{ab}$ on $\sigma$, the twist $\chi^{a}$ on $\sigma$, expansions $\theta_{(\ell)}$ and $\theta_{(k)}$ on $\sigma$, and the inaffinities $\kappa_{(\ell)}$ and $\kappa_{(k)}$ on $\sigma$~\cite{Ren90, Hay93EFE, BraDro95, Luk12}. 

We are not quite done yet, as we have not yet addressed the issue of existence of evolution of the initial data. Here we may take the approach of either the characteristic or Cauchy initial data problem. Rigorous theorems for the local existence and uniqueness of evolution of initial data often impose certain regularity conditions on the initial data. 

Consider first the standard (spacelike) Cauchy problem. Choquet-Bruhat's original 1952 theorem for (vacuum) Cauchy evolution~\cite{Cho52} required a triplet $(\Sigma, \gamma_{ab}, K_{ab})$, where $\Sigma$ is the initial (spacelike) Cauchy slice, $\gamma_{ab}$ its induced metric, and $K_{ab}$ its extrinsic curvature tensor, where $\gamma_{ab}$ is $C^{5}$ and $K_{ab}$ is $C^{4}$. Since then, these requirements have been progressively reduced to the requirement that the second partial weak derivatives of $g_{ab}$ be square integrable and of $K_{ab}$ once differentiable~\cite{HugKat77, ChoChr78,  ChoIse00, KlaRod02, Cho04, Max04, Max04b, SmiTat05}; the precise statement is that $g_{ab}$ is in the Sobolev space $H_{loc}^{3/2+\epsilon}$ and $K_{ab}$ in $H_{loc}^{1/2+\epsilon}$ for $\epsilon>0$. More recently, studies of low-regularity metrics in the context of junction conditions have produced limited proofs of local well-posedness for metrics with only \textit{first} partial weak derivatives being square integrable: $g_{ab}\in H_{loc}^{1}$ and $K_{ab}\in H_{loc}^{0}$~\cite{Cla97, VicWil01, GraMay08, SanVic15, SanVic16}. This is precisely the regularity regime of our desired results, and we will assume existence, in accordance with expectations partially borne out in this class of cases for the Cauchy problem.

For the characteristic problem (where only local existence and uniqueness results are rigorously established in broad generality, but see~\cite{CacFra04, CacNic06} for some limited global results), the original proof of Rendall for existence of a neighborhood at the intersection of two null hypersurfaces is given for $C^{\infty}$ data, but the expectation is that rougher initial data should behave similarly~\cite{Ren90}, while Hayward's proof in~\cite{Hay93EFE} shows that a unique solution exists up to caustics. More recently, Luk proved the existence of a neighborhood of the \textit{union} of both null hypersurfaces assuming the data is $C^{\infty}$~\cite{Luk12}, although followup work on impulsive gravitational waves in the characteristic problem has been able to accommodate a curvature with a Dirac $\delta$-function singularity in the vacuum~\cite{LukRod12}.

Note that uniqueness of the Cauchy evolution may be more questionable than existence, as it is possible that the initial  data could develop into a Cauchy horizon (this is not expected to occur for vacuum initial data~\cite{ChrPae12}). In that case, we simply adopt the approach in~\cite{Wald} and use the maximal Cauchy development.



Finally, a brief comment on the constraint equations for matter fields. The original approach of Rendall~\cite{Ren90} for proving (local) well-posedness of the characteristic initial data extends to scalar, Maxwell, and Yang-Mills fields coupled to gravity~\cite{ChrPae12}; a generalization of the method also works for Vlasov fields~\cite{ChoChr12}. This works well when the matter fields in $I_{W}[\sigma]$ and $O_{W}[\sigma]$ have the same matter Lagrangian.

\section{The Outer Entropy}\label{sec:outer}
We have thus far focused on giving a precise definition of minimar surfaces as generalizations of HRT surfaces. In what follows, we will give a definition of our generalization of the von Neumann entropy to the outer entropy. Like the outer wedge, the outer entropy is defined for any bulk surface homologous to the boundary. Although this is a purely classical bulk construction, its relation to the boundary entropy via the HRT formula will justify its interpretation as a coarse-grained entropy.

We are interested in the entropy associated to our ignorance of the inner wedge $I_{W}[\sigma]$ subject to knowledge of all of the field data (including the metric) in the outer wedge $O_{W}[\sigma]$. Consider all possible field data $\{\alpha\}$ for possible inner wedges $I_{W}^{(\alpha)}[\sigma]$ that could be patched onto $\sigma$ without altering $O_{W}[\sigma]$ (in such a way as to preserve all global conditions on the spacetime necessary to define the HRT/maximin surface). This is of course constrained by the matching conditions at $\sigma$ itself, derived in Sec.~\ref{sec:MultiJunction}. 

By AdS/CFT, each allowed spacetime obtained by some interior $I_{W}^{(\alpha)}[\sigma]$ then corresponds to some boundary state $\rho_{B}^{(\alpha)}$ whose von Neumann entropy is given by:
\be \label{eq:tomax}
S[\rho^{(\alpha)}] = -\mathrm{tr}(\rho_{B}^{(\alpha)} \ln \rho^{(\alpha)}) = \frac{\mathrm{Area}[X^{(\alpha)}]}{4 G\hbar},
\ee
where $X^{(\alpha)}$ is the HRT surface homologous to the boundary component $B$ in the spacetime with $I_{W}^{(\alpha)}[\mu]$. We would like to define an entropy associated with coarse graining over all such states $\rho_{B}^{(\alpha)}$. A simple way to do this is via a maximization of Eq.~\eqref{eq:tomax} over all states $\rho_{B}^{(\alpha)}$:

\paragraph{Outer Entropy:} The outer entropy associated to a surface $\sigma$ homologous to $B$  is defined by maximizing the von Neumann entropy over the possible inner wedge data $\{\alpha\}$:
\begin{equation}
S^{\text{(outer)}}[\sigma]\equiv \max\limits_{\{\alpha\}}[ -\mathrm{tr}(\rho_{B}^{(\alpha)} \ln \rho_{B}^{(\alpha)})].
\end{equation}

In other words, for any spacetime with $O_{W}[\sigma]$, we minimize the area of extremal surfaces homologous to $B$, and we then maximize over all possible inner wedges. This follows a familiar theme of min-max proposals for computing entropy in AdS/CFT~\cite{Wal12, FreHea16}.

\textit{A priori}, the outer entropy of a surface is not related to that surface's area. However, for an HRT surface $X_{B}$, $S^{(\mathrm{outer})}[X_{B}] = -\mathrm{tr}(\rho_{B}\ln \rho_{B})= \text{Area}[X_{B}]/4 G\hbar$. Our main result,  derived in Sec.~\ref{sec:main}, is an analogous relation for minimar surfaces $\mu$: $S^{(\mathrm{outer})}[\mu] = \text{Area}[\mu]/4 G\hbar$. On the other hand, for a large class of trapped and untrapped surfaces, we will show in Sec.~\ref{sec:general} that $S^{\text{(outer)}}$ is, respectively, larger and smaller than the area of the surfaces. This shows that minimar surfaces play a very special role in gravitational thermodynamics.


%
%
\section{Main Construction}\label{sec:main}
In this section, we prove that the outer entropy of a minimar surface is proportional to the area of that surface. We do this in three steps: first, we show that the outer entropy of a minimar surface $\mu$ is bounded from above by the area of $\mu$. This is done by showing that
\be \label{eq:inequality}
S[\rho_{B}^{(\alpha)}]=\frac{\mathrm{Area}[X_{B}^{(\alpha)}]}{4G\hbar}\leq \frac{\mathrm{\mathrm{Area}[\mu]} }{4G\hbar},
\ee
where as before $X_{B}^{(\alpha)}$ is the HRT surface of the connected component $B$ in the modified spacetime dual to $\rho^{(\alpha)}$. We will drop the superscript when the context is clear. The first equality follows by HRT, and the inequality will be shown via maximin techniques below.

Next, we show that for some choice of $\alpha$, there exists a spacetime $(M',g')$ with outer wedge $O_{W}[\mu]$ and an extremal surface $X$ whose area is exactly the same as the area of $\mu$.

Finally, we prove that $X$ is in fact the HRT surface of $(M',g')$. This constructs a spacetime dual to a boundary state $\rho'$ whose von Neumann entropy $S[\rho']$ is given by the area of $\mu$. We conclude that Eq.~\eqref{eq:inequality} is in fact an equality.

\subsection{Bounding the Outer Entropy}\label{sec:bound}
To show that $S^{(\mathrm{outer})}[\mu]$ is bounded from above by Area$[\mu]/4G \hbar$, we use a technique from~\cite{Wal12} of representing a marginal surface (e.g. $\mu$) on any Cauchy slice of the spacetime via a null congruence fired from $\mu$. We will assume for the rest of this section that  $\mu$ is marginally trapped ($\theta_{(\ell)}\leq 0$, $\theta_{(k)}=0$); the construction when $\mu$ is marginally anti-trapped ($\theta_{(\ell)}\geq 0$, $\theta_{(k)}=0$) is simply a time reverse.

\paragraph{Representative:} Let $\mu$ be a minimar surface in $M$, and let $\Sigma$ be a Cauchy slice of $M$, not necessarily containing $\mu$. Let $N_{k}(\mu)$ be the null congruence generated from $\mu$ by firing null geodesics in the $+k^{a}$ and $-k^{a}$ directions.  Since we are assuming global hyperbolicity, $N_{k}(\mu)$ splits $M$ into two pieces.  The representative of $\mu$ on $\Sigma$ can then be defined as
\be
\overline{\mu}(\Sigma) =N_{\pm k}(\mu)\cap \Sigma.
\ee
where by construction $\overline{\mu}(\Sigma)$ is homologous to $\mu$, $B$, and the HRT surface $X_B$.  (In some cases, when the generators of $N_{k}(\mu)$ all intersect prior to reaching $\Sigma$, $\overline{\mu}(\Sigma)$ may be the empty set.  In these cases, the HRT surface is also the empty set, and the following result will hold trivially.)

An immediate consequence of the NCC is that the area of $\tilde{\mu}(\Sigma)$ is bounded from above by the area of $\mu$~\cite{Wal12}:
\be \label{eq:bd}
\mathrm{Area}[\overline{\mu}(\Sigma)]\leq \mathrm{Area}[\mu].
\ee

Consider now any spacetime $(M^{(\alpha)},g^{(\alpha)})$ containing $O_{W}[\mu]$. As before, let $X_{B}^{(\alpha)}$ be the HRT surface of $B$.   By the maximin formulation \cite{Wal12} there exists a Cauchy slice $\Sigma$ of $(M^{(\alpha)},g^{(\alpha)})$ on which $X_{B}^{(\alpha)}$ is the minimal area surface homologous to $B$.  This immediately implies that the representative $\overline{\mu}[\Sigma]$ has greater area than that of $X_{B}$. Using Eq.~\ref{eq:bd} and HRT, we obtain
\be
S[\rho_{B}^{(\alpha)}] = \frac{\mathrm{Area}[X_{B}]}{4 G \hbar}\leq \frac{\mathrm{Area}[\overline{\mu}[\Sigma]]}{4 G \hbar} \leq  \frac{\mathrm{Area}[\mu]}{4 G \hbar}.
\ee
This establishes that for any spacetime $(M^{(\alpha)},g^{(\alpha)})$ with fixed outer wedge $O_{W}[\mu]$, the von Neumann entropy of $\rho_{B}^{(\alpha)}$ is bounded from above by one quarter of the area of $\mu$. Since $S^{\mathrm{(outer)}}[\mu]$ is obtained by maximizing $S$ subject to fixed $O_{W}[\mu]$, this immediately implies the desired result:
\be\label{eq:outerbound}
S^{\mathrm{(outer)}}[\mu] \leq \frac{\mathrm{Area}[\mu]}{4G\hbar}.
\ee

\subsection{Existence of Extremal Surface}\label{sec:exist}
We now proceed to give an explicit construction for the inner wedge $I_{W}[\mu]$ that maximizes $S[\rho_B]$.  In this new auxiliary spacetime $(M',g')$, we first show that there exists an extremal surface $X$ homologous to $B$ whose area is the same as the area of $\mu$.  (Later, in Sec.~\ref{subsec:min}, we will prove that $X$ is in fact the extremal surface of least area --- i.e. the HRT surface --- of $(M',g')$, so that the von Neumann entropy of the new spacetime is Area$[\mu]/4G\hbar$. This shows that the maximum of Eq.~\eqref{eq:inequality} is attained, and that Eq.~\ref{eq:outerbound} is in fact an equality.)

The spacetime is constructed via the initial data gluing procedure described in Sec.~\ref{sec:MultiJunction}.  The data in $O_{W}[\mu]$, and hence its past boundary $N_{-\ell}$, is already fixed.  To this (1) we glue a stationary null hypersurface $N_{-k}$ shot in the inwards-past direction from $\mu$; the assumption of stationarity fully fixes the data on $N_{-k}$.  We show (2) that there exists an extremal surface $X$ on $N_{-k}$ and calculate its location.  Finally, (3) we complete the spacetime on the other side of $X$ by assuming that it is CPT-reflection symmetric across $X$ (this introduces an additional AdS conformal boundary $\widetilde{B}$, and the reflection of the outer wedge $\widetilde{O}_{W}[\widetilde{\mu}]$ and its future boundary $\widetilde{N_{-\ell}}$.

\begin{figure}
	\begin{center}
	\includegraphics[width=0.5\textwidth]{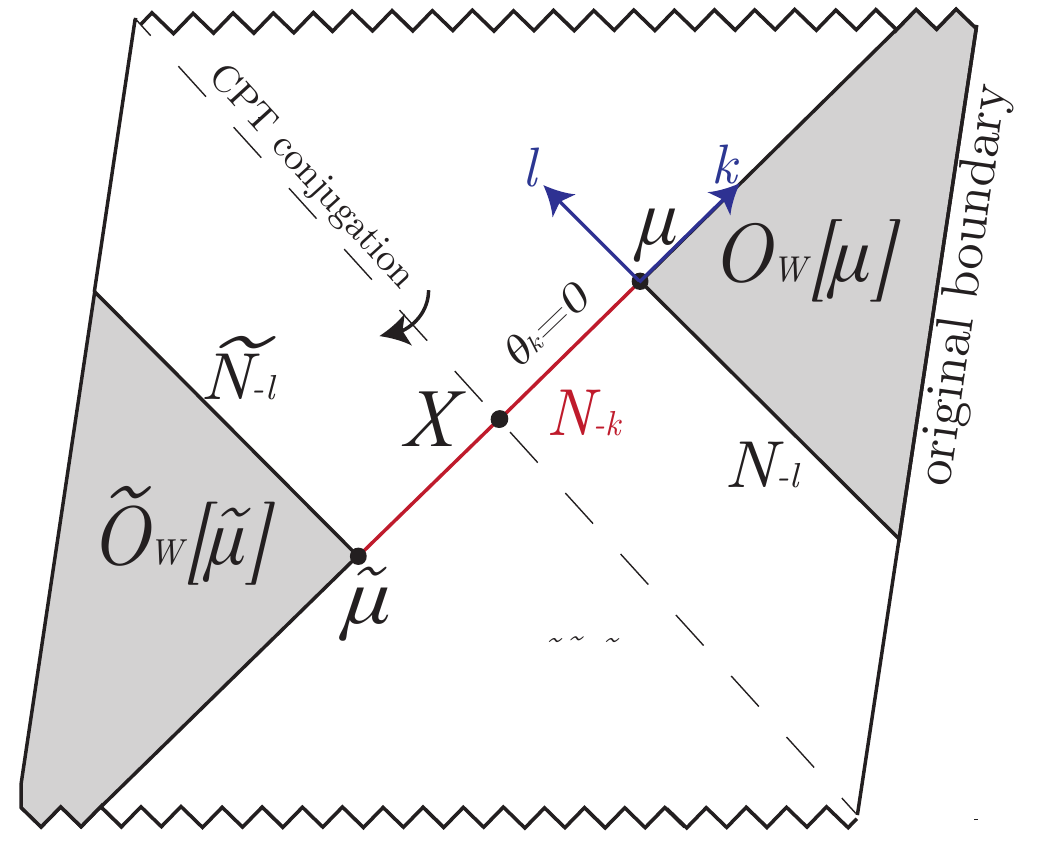}
	\end{center}
	\caption{A figure of the full maximizing spacetime, adapted from~\cite{EngWal17b}. The characteristic Cauchy slice that we construct to obtain this geometry consists of $N_{-\ell}\cup N_{-k}\cup \widetilde{N_\ell}$, }
\label{fig:maximizing}
\end{figure}

The geometry constructed so far includes a piecewise null Cauchy slice $\Sigma$ composed of three null hypersurfaces $N_{-\ell}$, $N_{-k}$, and $\widetilde{N_{-\ell}}$; this is illustrated in Fig.~\ref{fig:maximizing}.  The resulting hypersurface satisfies all constraint equations (and corresponding junction conditions) derived in Sec.~\ref{sec:MultiJunction}.  We can now define the entire spacetime $(M', g')$ by Cauchy evolution from this slice (the characteristic initial data problem).

Recall from Sec.~\ref{sec:MultiJunction} that in the characteristic problem, the data needed to be specified on $\Sigma$ to yield a well-defined and (at least locally) deterministic evolution is the following: the conformal metric on the null hypersurfaces, the null expansions of the null hypersurfaces at $\mu$, the twist, and the intrinsic metric on $\mu$.  All of this information is fixed by the  construction outlined above.

\subsubsection*{Step 1: Constructing $N_{-k}$}

Let us partially fix a gauge in null coordinates $u$ and $v$ and spatial coordinates $\{y_{i}\}_{i=1}^{d-1}$. Our choices below will not fully fix the gauge -- we still have some gauge freedom left in the spatial directions. The indices $\{a,b\}$ will run over all spacetime, while the indices $\{i,j\}$ are restricted to the transverse $y_{i}$ directions.

We fix $\mu$ to be at $u=0$ and $v=0$, so $N_{-k}$ is at $u=0$ and $N_{-\ell}$ is at $v=0$. In terms of the covariant definitions above, $\ell^{a}=(\partial/\partial u)^{a}$ and $k^{a}=(\partial/\partial v)^{a}$. We will use $\theta_{u}$ and $\theta_{v}$ below to emphasize our choice of gauge. Our gauge conditions are:
\begin{align} \label{eq:gauge}
 g_{uv}&=-1\\
 g_{uu}&=g_{ui}=0\\
 \left .  g_{vv} \right | _{u=0} &= \left . g_{vi}\right | _{u=0} = \left . g_{vv,u}\right |_{u=0}=0.
\end{align}
In this gauge, the twist and null extrinsic curvatures are:
\begin{align}
& g_{vi,u}= 2\chi_{i}\\
& \left . g_{ij,u}\right |_{v=\mathrm{const}} = 2B_{ij}\phantom{}_{(u)}\\
& \left . g_{ij,v}\right |_{u=0}=2 B_{ij}\phantom{}_{(v)}
\end{align}
\noindent where $\chi_{i}$ is the (gauge-dependent) twist and $B_{ij(q)}$ is the null extrinsic curvature in the $q=u$ or $q=v$ direction, which is related to the expansion and shear via Eq.~\eqref{eq:Bdecomp}.

In this gauge, the constraint equations reduce to (see e.g.~\cite{Hay01, Hay04, GouJar06, Hay06, Cao10} for the gauge-independent constraints):
\begin{align} &\theta_{u,v}=-\frac{1}{2} \mathcal{R} +\nabla\cdot \chi -\theta_{u}\theta_{v} + 8\pi G T_{uv}+\chi^{2}  \label{eq:GUV} ,\\
& \theta_{v,v} =- \frac{1}{D-2}\theta_{v}^{2} -\varsigma_{v}^{2} -8\pi G_{N} T_{vv}, \label{eq:Raych}\\
&\chi_{i,v}=  -\theta_{v}\chi_{i} +  \left(\frac{D-3}{D-2} \right)\nabla_{i} \theta_{v} - (\nabla\cdot \varsigma_{v})_{i} +8 \pi T_{iv}
\end{align}
where $\theta_{u,v}$ is the $v$ derivative of $\theta_{u}$ for constant $v$ slices and $T_{vv}$ denotes the $v-v$ component of $T_{ab}$. 

We now specify initial data on $N_{-k}$. Because we are fixing $O_{W}[\mu]$ and will implement a symmetry transformation to complete the spacetime, this is the only hypersurface on which we need to specify initial data. We will require:
\begin{align} 
& \varsigma_{v}[N_{-k}]=0, \label{shear} \\
& T_{vv}[N_{-k}]=0, \label{focus}\\
&T_{iv}[N_{-k}] = 0, \label{twist} \\
&T_{uv}[N_{-k}]= \text{const}.  \label{cross}
\end{align}
Inserting the first two equalities into the Raychaudhuri Eq. \eqref{eq:Raych} implies that $\theta_{v}=0$; hence $N_{-k}$ is stationary, and $\mathcal{R}$ is also constant along $N_{-k}$.  Eq. \eqref{twist} is a condition on the twist, since on $N_{-k}$, $T_{iv}=\chi_{i,v}$; this fixes the twist to be constant along the $v$-direction on $N_{-k}$. Using the above, Eq. \eqref{cross} requires $\left . \theta_{u,v}\right |_{v=\text{const}}$  to be a constant along the $v$-direction via Eq.~\eqref{eq:GUV}. 

The reader may ask whether we can always choose the stress tensor to obey our requirements. Let us briefly justify this in the special case where the bulk matter is a scalar field $\phi$ (with the standard kinetic action) minimally coupled to a Maxwell field $A_{a}$.  The Lagrangian density is
\begin{equation}
{\cal L} = -\sqrt{-g} \left (  \frac{1}{4} F_{ab}F_{cd}\, g^{ac}g^{bd}+ \bar{\nabla}_{a}\phi^* \bar{\nabla}_{b}\phi \, g^{ab} \right),
\end{equation}
where $\bar{\nabla}_{a}$ is the covariant derivative with respect to the gauge potential $A_{a}$.
With the corresponding stress-energy tensor:
\begin{align}
& T_{vv}=2\bar{\nabla}_{v}\phi^*\bar{\nabla}_{v}\phi + F_{vi}F_{v}\,^{i}, \\
&T_{vi}=2\bar{\nabla}_{v}\phi^*\bar{\nabla}_{i}\phi + F_{vj}F_{i}\,^{j} + F_{vi}F_{uv},\\
& T_{uv} = \bar{\nabla}_{i}\phi^* \bar{\nabla}_{i}\phi +  F_{ij}F^{ij} + \frac{1}{2}F_{uv}F_{uv}. \label{eq:Tuv}
\end{align}

By setting $ \bar{\nabla}_{v}\phi = F_{iv} = 0$ on $N_{-k}$, these being free data in the characteristic problem, we immediately recover Eqs.~\eqref{focus}-\eqref{cross}. (To prove constancy of $T_{uv}$,  use  the Bianchi identity and the Gauss Law on the null hypersurface.) These conditions are analogous to Eq.~\eqref{shear} for gravitational radiation. 

We expect that a similar prescription exists for other reasonable matter fields to satisfy Eqs.~\eqref{focus}-\eqref{cross}. Assuming so, it is always possible to construct a stationary hypersurface null $N_{-k}$ satisfying the constraint equations.  

We now show that we can also satisfy the junction conditions Eq.~\eqref{contThK}-\eqref{contChi} across $\mu$, as well as the corresponding continuity conditions for the matter fields:\footnote{When $\mu$ is not simply connected, we should also demand matching of the integral of $A_i$ along noncontractible Wilson loops.}
\bea 
& [\phi] =0, \label{m1}\\
& [F_{ij}]=0,  \label{m2}\\
& [F_{uv}]=0  \label{m3}.
\eea

The first junction Eq. ~\eqref{contThK} for $\theta_v$ is already satisfied because it vanishes on both $N_{-k}$ and $\mu$.  To see that we can satisfy the remaining conditions, note that so far we have only fixed the transverse geometry $g_{ij}$, the twist $\chi_{i}$ and $\left. \theta_{u,v}[N_{-k}]\right|_{v=\text{const}}$ up to functions of the transverse $y_{i}$ directions.  Even after fixing $\theta_{u,v}$ we can still choose $\theta_{u}$ at $v = 0$ on $N_{-k}$.  Similarly, $\phi$ and $F_{ij}$ are defined up to transverse functions.  We are therefore free to choose all of these quantities to be continuous across the junction at $\mu$:
\begin{align} 
& g_{ij}[N_{-k}]=g_{ij}[\mu],\\
&\chi_{i}[N_{-k}]=\chi_{i}[\mu], \\
&\left. \theta_{u,v}[N_{-k}]\right|_{v=\text{const}}= \theta_{u,v}[\mu], \\
&\lim_{v\rightarrow 0}\left. \theta_{u}[N_{-k}]\right|_{v=\text{const}}= \theta_{u}[\mu],\\
& \phi[N_{-k}]=\phi[\mu],\\
& F_{ij}[N_{-k}]=F_{ij}[\mu],\\
& F_{uv}[N_{-k}]=F_{uv}[\mu],
\end{align}
where the last three conditions also imply $T_{uv}[N_{-k}]=T_{uv}[\mu]$.  Note that the cross-focusing constraint \eqref{eq:GUV} continues to be satisfied because all of its terms are by construction the same on $N_{-k}$ and $\mu$; this is only possible because $\theta_{v}=0$ on $\mu$.

These choices satisfy the junction conditions for the metric Eqs.~\eqref{contThL}\&\eqref{contChi}, as well as for matter Eqs.~\eqref{m1}-\eqref{m3}.  We have therefore succeeded at our goal of gluing a stationary null hypersurface $N_{-k}$ to $\mu$.\footnote{The shear $\varsigma_{v}$ may be discontinuous across $\mu$, but 
such solutions are believed to be valid characteristic initial data \cite{Ren90}; indeed, Ref.~\cite{LukRod12, LukRod13} studied a discontinuous shear sourcing a $\delta$-function in the curvature, which was still distributionally well-behaved. (We expect that this continues to be true even if the shear discontinuity reflects off of the AdS boundary.)} 






\subsubsection*{Step 2: Finding $X$}

 Let us now proceed to find  an extremal surface on $N_{-k}$. By construction, $\theta_{(k)}[N_{-k}]=0$. Finding an extremal surface $X$ therefore reduces to finding a cross-section of $N_{-k}$ on which $\theta_{(\ell)}=0$.  On a constant $v$ slice, all of the terms on the RHS of Eq.~\eqref{eq:GUV} --- including the contributions to $T_{uv}$ from Eq.~\eqref{eq:Tuv} --- are constant, and equal to their values at $\mu$. By definition of a minimar surface (Requirement \ref{def:minimarCross}) $\theta_{u,v}[\mu]<0$, so $\left. \theta_{u,v}\right |_{v=\text{const}}$ is strictly negative and constant with respect to $v$.
Hence $\left. \theta_{u}\right |_{v=\text{const}}$ increases at a constant rate as we move to more negative $v$ values. Since $\left. \theta_{u}\right |_{v=\text{const}}$ starts out negative (by Eq. \eqref{def:marNeg}), this indicates that it attains zero on some slice $\sigma$ of $N_{-k}$. However, this need not be a constant-$v$ slice, and if not, then generally $\left. \theta_{u}\right |_\sigma \ne \left. \theta_{u}\right |_{v=\text{const}}$, since (as described explicitly below) $\theta_{u}$ is also sensitive to the derivatives of $v$ with respect to the transverse directions.  We must therefore work harder to find the slice $X$ on which $\theta_{u}$ vanishes; this slice will have two vanishing null expansions, and thus by definition it would be an extremal surface.

To find this putative slice, we first compare $\theta_{u}$ of a varying-$v$ slice $\beta$ with that of a constant-$v$ slice $\alpha$. Let $v=f(y^{i})$ be the equation for the location of $\beta$ on $N_{-k}$.

\begin{figure}
\centering
\includegraphics[width=0.5\textwidth]{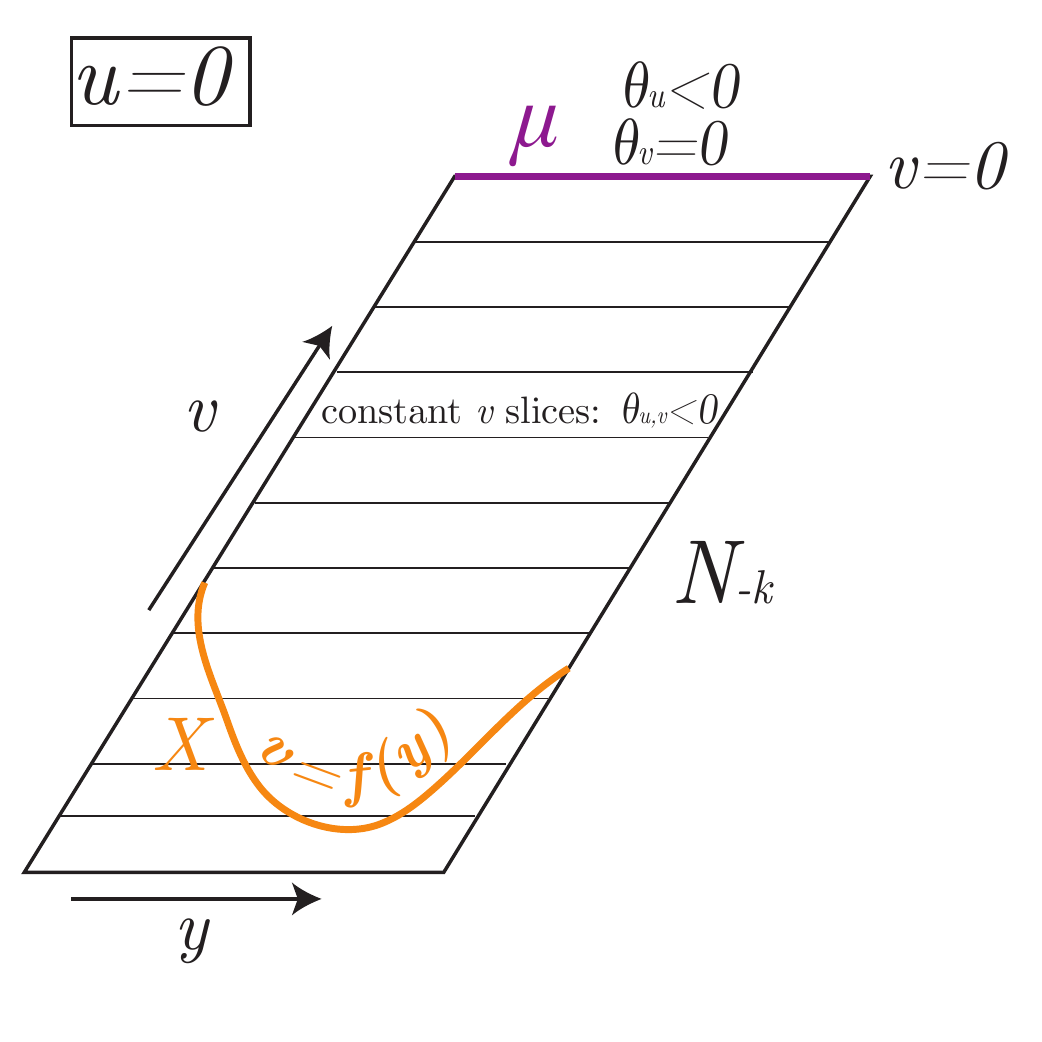}
\caption{A (1+1)-dimensional cartoon illustrating the hypersurface $N_{-k}$, which is located at $u=0$. The minimar surface $\mu$ is at $v=0=u$. The horizontal black lines are slice of constant $v$, on which $\theta_{u,v}<0$, and $v<0$ on the entire drawn slice. The extremal surface $X$ is in general not a contant $v$ slice, but rather some function $v=f(y^{i})$, drawn above in orange.}
\label{fig:affineslices}
\end{figure}

By definition, $v^{a}$ is normal to both $\alpha$ and $\beta$, but $u^a$ is normal only to $\alpha$.  The second null normal to $\beta$, denoted $w^{a}$, can be computed from the defining equation for $\beta$:
\be
w^{a} = u^{a} + \sum\limits_{i} y_{i}^{a}\nabla_{i}f + \frac{1}{2}v^{a}\Box f ,
\ee
where $\nabla_{i}\equiv  y^{b}_{i}\nabla_{b}$. The null expansion of $\beta$ in the $w^{a}$ direction is given by $\theta_{w}[\beta]=h^{ab}\nabla_{a}w^{b}$, where $h_{ab}$ is the induced metric on $\beta$. Transforming this into the coordinates on $\alpha$ yields the following relation:
\be \label{eq:relation}
\theta_{w}[\beta]= \theta_{u}[\alpha]+\Box  f(y_{i}) + 2\chi\cdot \nabla  f(y_{i}),
\ee
where we have used stationarity of $N_{-k}$ to simplify the equation.   This is illustrated in Fig.~\ref{fig:affineslices}.

Because $\theta_{u,v}$ (i.e. $\partial_{v}\theta_{u}$)  is independent of $v$ on constant-$v$ slices, $\theta_{u}$ of $\alpha$ can be simply written in terms of the expansion at $\mu$:
\be\label{eq:affineexp}
\theta_{u}[\alpha] = \theta_{u}[\mu]+v\,\theta_{u,v}[\mu].
\ee
Thus we obtain
\be \label{eq:loc}
\theta_{w}[\beta]= \theta_{u}[\mu] +\theta_{u,v}[\mu] f(y_{i}) +\Box f(y_{i}) + 2\chi\cdot \nabla f(y_{i})\equiv L^{\mu}[f] + \theta_{u}[\mu],
\ee
where $L^{\mu}$ is an operator that depends only on quantities evaluated on the minimar surface $\mu$. This operator is known as the \textit{stability operator}~\cite{AndMar05, AndMar08}.

Recall now that we are searching for an extremal surface: that is, we are looking for a surface, which we shall call $X$, where $\theta_{w}[X]$ vanishes. Eq.~\ref{eq:loc} becomes the following eigenvalue equation:
\be
L^{\mu}[f] = -\theta_{u}[\mu]
\ee

It is a known result that the eigenvalue of $L[f]$ with the smallest real value is real; furthermore, if and only if the marginal surface $\mu$ is ``strictly stable'' (equivalent to Requirement \ref{def:minimarCross} for a minimar surface)  this eigenvalue is strictly larger than zero~\cite{AndMar05}. Hence, because $\mu$ is minimar,  $L[f]$ has no zero eigenvalues, and is thus invertible. A nontrivial solution exists, and therefore the sought-after extremal surface $X$ exists and may be found by solving Eq.~\eqref{eq:loc}.

A final property that we will need is that $f[y]$ is negative  (or zero): otherwise, it could lie partly on $N_{+k}$;  in such a case, we would find that we need to replace data on $O_{W}[\mu]$ to get an extremal surface with the area of $\mu$, which would ruin our construction. 

That $f$ must be nonpositive can be shown from stability of $\mu$ by invoking the Krein-Rutman theorem~\cite{ KreRutEnglish, KreRutRussian, AndMar08}, but a simple geometric argument proves this result as well. Suppose that $f$ is indeed positive somewhere, so that it lies at least partly on $N_{+k}$. Because $X$ is compact, $f$ has a maximum. At the maximum, $\nabla_{i} f=0$, $\Box f\leq0$, and $f>0$. But in such a case, Eq.~\eqref{eq:loc} for an extremal $\beta$ has a strictly negative quantity on the right hand side but is zero on the left hand side. So we have a contradiction, and thus $f \le 0$.

We conclude that there exists an extremal surface $X$ on $N_{-k}$. Because $N_{-k}$ is stationary, Area$[X]=\text{Area}[\mu]$.

\subsubsection*{Step 3: CPT Reflecting}

So far we have constructed initial data on a partial Cauchy slice $N_{-k} \cup N_{-\ell}$ terminating on $X$ in the interior.  Assuming that $X$ is indeed an HRT surface, then by evolving from this initial data (using the AdS boundary conditions) we expect to be able to construct the entanglement wedge $E_W[B]=O_W[X]$, which is dual to some mixed state $\rho_{B}'$ of the boundary CFT associated with $B$. 

The advantage of this is that it gives us a state associated with a single spacetime boundary region $B$, which is natural if our original spacetime $M$ had only a single boundary CFT (e.g. in the case of a collapsing black hole).  However, this leads to the oddity of a bulk spacetime which simply ends at $X$.  In order to construct a complete spacetime $M'$, and to facilitate our proof that $X$ is indeed an HRT surface, we will find it convenient to construct a spacetime with an additional auxiliary boundary $\widetilde{B}$.  In the CFT dual, this corresponds to adding a second CFT that purifies the state, similar to the thermofield double interpretation of the Einstein-Rosen wormhole geometry \cite{Mal01}.

Accordingly, to complete our construction, we must specify initial data on a complete Cauchy slice.  We will generate this slice by acting with a CPT-reflection across $X$ that takes $v \to -v$, $u \to -u$, $y_i \to y_i$.  This transformation acts on the initial data at a surface as follows:

\begin{table}[h!]
\begin{center}
  \begin{tabular}{  c | c}

    \textbf{CPT Odd} & \textbf{CPT Even} \\ 
    \hline \hline
   $ \theta_{v}$ & $\chi_{i}$\\
$\theta_{u}$ & $ \phi $   \\
$F_{iv}$ & $F_{ij}$ \\
\ & $F_{uv}$ 
  \end{tabular}
  \end{center}
\end{table}
All quantities that are odd under CPT vanish on $X$ by construction.\footnote{$F_{iu}$ is excluded from the table because it is not needed as initial data for the characteristic initial data formulation of electromagnetism. Furthermore,  continuity of $F_{iu}$ and $F_{iv}$ is not required as junction conditions.}  Therefore the CPT-conjugate data satisfies the requisite junction conditions Eqs.~\eqref{contThK}-\eqref{contChi} and \eqref{m1}-\eqref{m3}.\footnote{Note that the full CPT is required for this, since the twist $\chi_i$ is odd under $P$ and $T$ separately, while the gauge potential $A_a$ has an extra sign in its transformation under $C$ and $T$ separately.}

The result is a second boundary $\widetilde{B}$, with time running in the opposite direction from $B$. $B$ and $\tilde{B}$ are connected by a Cauchy slice with three null segments: $\Sigma=N_{-\ell}[\mu]\cup N_{-k}[\mu]\cup \widetilde{N_{-\ell}}[\widetilde{\mu}]$. We are using tildes to represent CPT-conjugated submanifolds. This is illustrated in Fig.~\ref{fig:maximizing}.\footnote{We believe that the resulting spacetime is the bulk dual of the GNS construction \cite{GelNeu43, Seg47} acting on the state $\rho_{B}'$ in the algebra of $B$. The GNS construction is a natural purification of the state respecting all symmetries, including a $\mathbb{Z}_2$ antiunitary symmetry relating $B$ to a complementary system $\widetilde{B}$.}

We have now specified all data necessary to uniquely evolve characteristic initial data via the Einstein field equation.  The resulting spacetime $(M',g')$ has a minimar surface $\mu$ whose outer wedge $O_{W}[\mu]$ is the same as in $(M,g)$. $M'$ contains an extremal surface $X$ on the boundary of the inner wedge $I_{W}[\mu]$, which is homologous to $\mu$ and therefore to the boundary $B$.  Thus it is a candidate for the HRT surface, although we have not yet shown it is the extremal surface of least area in $M'$.  That will be accomplished in the next section.

%
%
%
%
%
%
\subsection{Minimality of the Extremal Surface}\label{subsec:min}
We now proceed first to show that the von Neumann entropy of $(M',g')$ is actually given by Area$[\mu] = \text{Area}[X]$. This amounts to showing that the area of any other extremal surface $X'$ in $(M',g')$ cannot exceed the area of $X$.  If there are no other extremal surfaces in $(M',g')$, we are done. So suppose that there exists an extremal surface $X'\neq X$ homologous to $B$.

Let $\Sigma$ be the Cauchy slice $\Sigma_\text{min} \cup N_{-k}\cup\widetilde{\Sigma_\text{min}}$ of $(M',g')$. Recall that $\Sigma_\text{min}[\mu]$ is the partial Cauchy slice on which the minimar surface $\mu$ is minimal (Requirement~\ref{def:minimarMin}), and  $\widetilde{\Sigma_\text{min}}[\widetilde{\mu}]$ is its CPT-conjugate.

\begin{figure}[t]
\centering
\subfigure[]{
\includegraphics[width=0.4\textwidth]{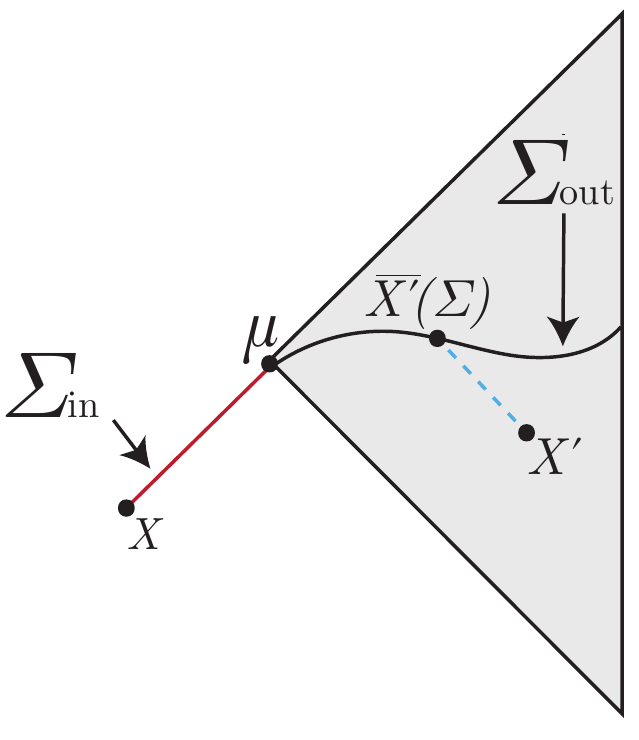}
\label{subfig:MinExt}
}
\hspace{0.5cm}
\subfigure[]{
\includegraphics[width=0.4\textwidth]{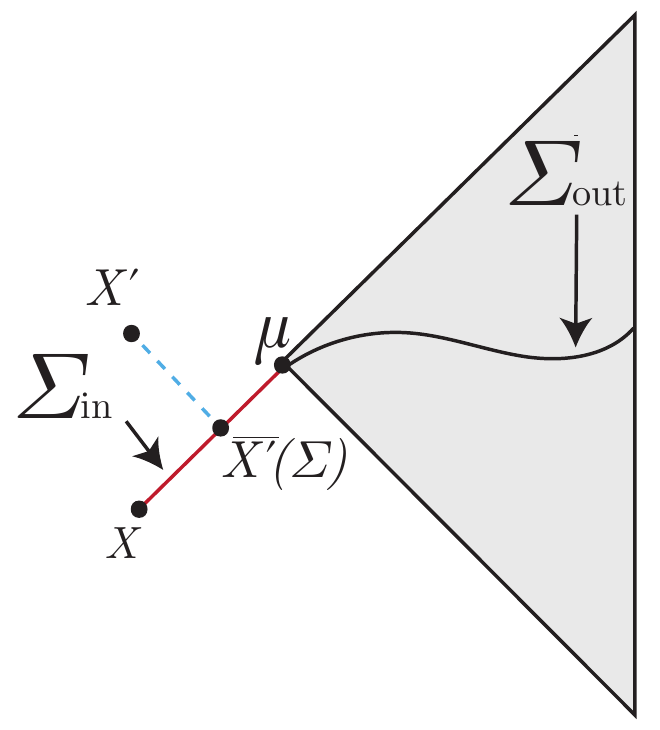}
\label{subfig:MinInt}
}
\caption{(a) The case where there is an extremal surface $X'$ in $O_{W}[\mu]$. The representative $\overline{X'}(\Sigma)$ lies on $\Sigma_{\text{out}}$, and its area is smaller than (or equal to) that of $X$, but by minimality of $\mu$ on $\Sigma$, this area must still be larger than that of $\mu$. (b) If $X'$ has a representative on $\Sigma_{\text{in}}= N_{-k}$, then the area of the representative is equal to the area of $\mu$ by stationarity of $N_{-k}$ and again no larger than the area of $X'$.}
\label{fig:MinExtorInt}
\end{figure}

Let $\overline{X'}(\Sigma)$ be the representative of $X'$ on $\Sigma$. We will first treat the case where $\overline{X'}(\Sigma)$ lies on just one of $\Sigma_\text{min}$ or $ N_{-k}$ (the $\widetilde{\Sigma_\text{min}}$ case is symmetrical). This is illustrated in Fig.~\ref{fig:MinExtorInt}.

If $\overline{X'}(\Sigma)$ lies on $ N_{-k}$, then:
\be
\mathrm{Area}[\overline{X'}(\Sigma)]=\mathrm{Area}[X].
\ee
This follows from the fact that $N_{-k}$ is stationary.

If $\overline{X'}(\Sigma)$ lies on  $\Sigma_\text{min}$, then by definition of $\mu$, the area of $\overline{X'}(\Sigma)$ must be larger than the area of $\mu$. Altogether, if $\overline{X'}(\Sigma)$ lies on either side of $\mu$, we have:
\be
\mathrm{Area}[\overline{X'}(\Sigma)]\geq\mathrm{Area}[X].
\ee
Since the area of a representative of an extremal surface is always larger than the area of the extremal surface, we immediately find:
\be
\mathrm{Area}[X']\geq\mathrm{Area}[X].
\ee
This shows that $X$ is the minimal area extremal surface homologous to $B$. It is possible $X'$ has the same area, but this does not affect our conclusion, that the area of $\mu$ gives the von Neumann entropy of $(M', g')$.

The case where $\overline{X'}(\Sigma)$ intersects multiple regions is only slightly more complicated. Suppose for example that $\overline{X'}(\Sigma)$ lies on both $N_{-k}$ and $\Sigma_\text{min}$. Let
\begin{align}
&x_{1}=\overline{X'}\cap \Sigma_\text{min}\\
&x_{2}=\overline{X'}\cap N_{-k},
\end{align}
and further divide $\mu$ into two subsets, where $\mu_{1}=\mu\cap O_{W}[X']$ and $\mu_{\mathrm{2}}$ is the complement in $\mu$. See Fig.~\ref{fig:mixed}. Note that $\mu_{1}\cup x_{2}$ and $\mu_{2}\cup x_{1}$ are both homologous to $\mu$. Then we find:
\begin{figure}
\centering
\includegraphics[width=0.5\textwidth]{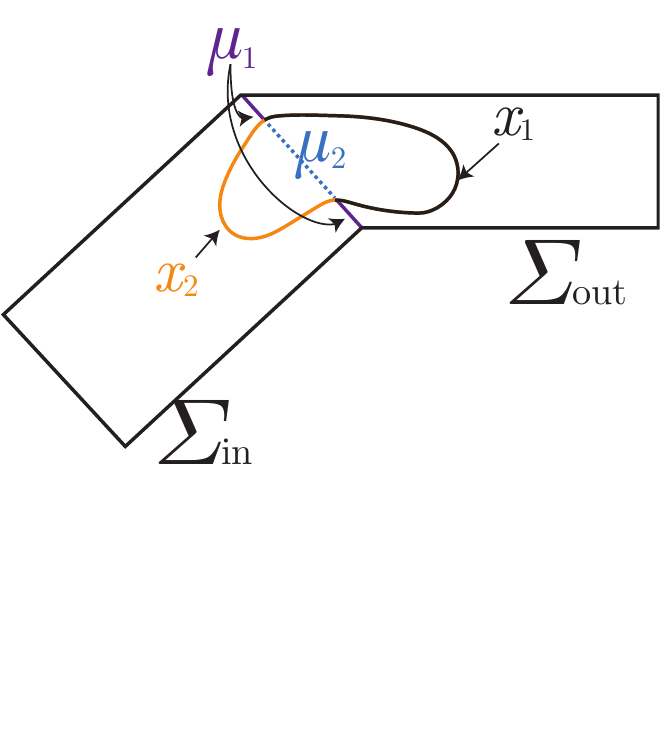}
\caption{The case of a surface $X'$ whose representative lies partly in $\Sigma_{\text{out}}$ and partly in $\Sigma_{\text{in}}$. This illustration is a planar projection of the spacetime and should be thought of as being periodically identified. }
\label{fig:mixed}
\end{figure}
\begin{align}
\mathrm{Area}[\mu_1] +\mathrm{Area}[x_{1}]&\geq\mathrm{Area}[\mu]\\
\mathrm{Area}[\mu_{2}]+\mathrm{Area}[x_{2}]& = \mathrm{Area}[\mu],
\end{align}
where the first line follows from Requirement~\ref{def:minimarMin}, and the second follows by stationarity of $N_{-k}$. Altogether, we find:
\be
\mathrm{Area}[X']\geq \overline{X'}(\Sigma)= \mathrm{Area}[x_{1}]+\mathrm{Area}[x_{2}] \geq \mathrm{Area}[\mu].
\ee
(If $\overline{X'}(\Sigma)$ intersects all three components of $\Sigma$, the argument works the same way.) This completes the proof.

This shows that $X$ is minimal among all extremal surfaces in the auxiliary spacetime $(M',g')$ and therefore by HRT~\eqref{eq:HRT}:
\be S[\rho']=\frac{\mathrm{Area}[\mu]}{4G\hbar},
\ee
where $\rho'$ is the state dual to $(M',g')$.

We already established in Sec.~\ref{sec:bound} that the outer entropy is bounded above by $\mathrm{Area}[\mu]/{4G\hbar}$.  This construction shows that the maximum can indeed be attained. This proves the desired claim:
\be
S^{\mathrm{(outer)}}[\mu] = \max\limits_{\{\alpha\}}[S_{vN}[\rho_{B}^{(\alpha)}] = \frac{\mathrm{Area}[\mu]}{4G\hbar}.
\ee


\section{Outer Entropy of Other Surfaces}\label{sec:others}

\subsection{Extremal Surfaces as Minimar Surfaces}\label{sec:extremal}

A special case of our result is when the minimar surface is itself an extremal surface $X$, so that $\theta_{(\ell)} = 0$ as well as $\theta_{(k)} = 0$.  To be minimar, the surface must either be HRT already (in which case $S^{(\text{outer})} = S_{vN}$ trivially), or else a non-minimal extremal surface lying closer to the boundary than any other extremal surface of lesser area.  (As shown in Appendix \ref{sec:HRT}, in the extremal case the minimality on the partial Cauchy slice $\Sigma_\text{min}$  automatically implies that $\theta_{(\ell),k} \le 0$.)  

In this case, there is no need to construct $N_{-k}$.  We only need to perform the CPT-reflection about $X$, which is now guaranteed by  Sec.~\ref{subsec:min} to be the HRT surface in the new spacetime.  Hence 
\be
S^{(\text{outer})} = \frac{\text{Area}[X]}{4G_{N}\hbar}
\ee
allowing us to interpret the area of a non-minimal extremal surface as a coarse-grained entropy.

This construction works even if we take $X$ to be an extremal surface anchored to the boundary of a subregion $R \in B$.  Because $X$ and the original HRT surface are locally minimal on some Cauchy slice $\Sigma$ of the original spacetime $(M,g)$, it follows that the divergence structure of their areas agrees.

However, in general, as investigated in~\cite{MarWhi17} the divergences in the area of general boundary-anchored marginal surfaces are local to the boundary region in question only at leading order. This complicates any attempt to giving a similar prescription for the outer entropy of a non-extremal minimar surface.  Hence we do not address the general case of boundary-anchored $\mu$'s in this paper.

\subsection{Non-Marginal Surfaces}\label{sec:general}

How does the outer entropy of non-minimar surfaces compare with their area? There are no general grounds to expect a particular relationship between the area and outer entropy of arbitrary surfaces. However, for untrapped and certain trapped surfaces, the area turns out to be a bound on the outer entropy.

\subsubsection{Untrapped Surfaces}\label{sec:untrapped}

Recall that an untrapped surface satisfies the following relation:
\begin{align}
&\theta_{(\ell)}<0\\
&\theta_{(k)}>0.
\end{align}
An example of such a surface is a cross-section of a generic causal horizons, for which $\theta_{(k)} > 0$ due to Hawking's area-increase theorem \cite{Haw71} while $\theta_{(\ell)}< 0$ if the cross-section is outside the past horizon.  An even more special case is a (generic) casual surface, which is the intersection of the past and future causal horizons \cite{HubRan12}.  It has been proposed that the areas of these surfaces correspond to some notion of coarse-grained entropy \cite{HubRan12,KelWal13}, with specific proposals being given in \cite{FreMos13,KelWal13}.  However, the proposal in \cite{FreMos13} was refuted by \cite{KelWal13}, while \cite{KelWal13} was refuted in \cite{EngWal17a}.

The counterexample in \cite{EngWal17a} involved a causal surface with $S^{\mathrm{outer}} = 0$ but $\text{Area} >0$.  Hence the outer entropy was strictly greater than the area.  Below we prove that this relationship in fact holds more generally for any untrapped surface that lives outside of the horizons\footnote{A similar conclusion is reached in~\cite{NomRem18}.}:
 
\paragraph{An Upper Bound on $S^{(\mathrm{outer})}$:} If $\upsilon$ is an untrapped surface homologous to $B$ and lying outside the past and future horizons of $\partial M$, then:
\begin{equation}\label{untrapped}
S^{(\mathrm{outer})}[\upsilon] < \frac{\mathrm{Area}[\upsilon]}{4G\hbar}.
\end{equation}

\begin{proof}
Because $\upsilon$ lies outside of the event horizons, a nontrivial compact extremal surface can only live in $I_{W}[\upsilon]$ \cite{Pen69}. Let $X$ be the HRT surface corresponding to a given choice of $I_{W}[\upsilon]$, and let $\Sigma_{\text{min}}$ be the Cauchy slice on which $X$ is minimal. We find:
\be
\text{Area}[X]\leq \text{Area}[\Sigma_{\text{min}}\cap \partial I_{W}[\upsilon]]\leq \mathrm{Area}[\upsilon],
\ee
where the first inequality follows by minimality of $X$ on $\Sigma_{\text{min}}$, and the second inequality follows from the fact that $\upsilon$ is untrapped, so that the area decreases as we move inwards on the null hypersurface $\partial I_{W}$. 

The equality in the first equation can only happen if there are multiple minima on $\Sigma_{\text{min}}$ and $\Sigma_{\text{min}}\cap \partial I_{W}[\upsilon]$ is another of those minima; the equality in the second equation can only happen if $\upsilon$ lies on $\Sigma_{\text{min}}$ since otherwise there is some focusing.  But $\upsilon$ cannot be a minimum area surface on any Cauchy slice, since the outward spacelike expansion on $\Sigma$ (which is a linear combination of $\theta_{(k)}$ and $-\theta_{(\ell)}$) is nonzero.  Hence the inequality \eqref{untrapped} is strict.

\end{proof}

The situation is more complicated for untrapped surfaces inside an event horizon.  But it is likely that a maximin argument \cite{Wal12} can be made towards the same conclusion.~\footnote{The argument would roughly work by constructing a surface which is maximin in $I_{W}[\upsilon]$, and showing that its area is smaller than that of $\upsilon$; the conclusion then follows immediately. A possible issue, however, is the equivalence of maximin and HRT if the maximin surface lies on $\partial I_{W}[\mu]$; a complete proof would likely require showing that the maximin surface cannot do so.}

\subsubsection{Trapped Surfaces}\label{sec:trapped}
An opposite bound can be proven for a class of trapped surfaces, i.e. surfaces with:
\begin{align}
&\theta_{(\ell)}<0\\
&\theta_{(k)}>0.
\end{align}
Here we wish to show that $S^{\text(outer)}$ always exceeds the area, but we need some additional assumption to rule out cases where the trapped surface lies in a ``bag of gold'' region \cite{Whe64, Mar08} behind another black hole with small area.  For this reason, we require our trapped surface to be to the outward-null future of a minimar surface:

\paragraph{Lower Bound on $S^{\mathrm{(outer)}}$:} Let $\mu$ be a minimar surface (with $\theta_{(\ell)}<0$) and let $\tau$ be a trapped surface on the null congruence $N_{+k}$ fired in the $+k^{a}$ direction from $\mu$. Then:
\begin{equation}
S^{(\mathrm{outer})}[\tau] \geq \frac{\mathrm{Area}[\upsilon]}{4G\hbar}.
\end{equation}

\begin{proof} This is almost immediate from the definition of $\tau$:
\be
\frac{\mathrm{Area}[\tau]}{4G\hbar}< \frac{\mathrm{Area}[\mu]}{4G\hbar} = S^{\mathrm{(outer)}}[\mu] \leq S^{\mathrm{(outer)}}[\tau],
\ee
\noindent where the first inequality follows by focusing, the equality follows from the fact that $\mu$ is minimar; the second inequality follows from $O_{W}[\tau]\subset O_{W}[\mu]$, so to obtain $S^{\mathrm{(outer)}}[\tau]$ we must coarse grain over fewer constraints than we do to obtain $S^{\mathrm{(outer)}}[\mu]$. Therefore the former must be at least as large as the latter. This establishes that for all trapped surfaces on $\partial O_{W}[\mu]$ for a minimar surface $\mu$,  the area gives an upper bound on the outer entropy.  The construction is similar for anti-trapped surfaces ($\theta_{(\ell)}>0$).
\end{proof}

We expect, however, that $S^{(\mathrm{outer})}[\tau]$ cannot be made arbitrarily large, since $O_W[\tau]$ includes a boundary slice $B$, and in AdS/CFT we expect that there is a maximum entropy state compatible with a finite ADM mass.

\section{Boundary Perspective: The Simple Entropy}\label{sec:simple}

Our focus has thus far been on proving that the outer entropy of a minimar surface $\mu$ --- the entropy associated with coarse graining over inner wedge of $\mu$ subject to knowing its outer wedge --- is proportional to the area of $\mu$.  Aside from the use of HRT to interpret the area of the extremal surface $X$ as $S_{vN}$, this statement has been defined entirely on the \emph{bulk} side of the duality.
Yet to get a fully holographic definition of the coarse-grained entropy of a black hole, we need to define the dual quantity on the \emph{boundary} side, using as little of the bulk physics as possible.


We therefore give a proposal for the boundary interpretation, \textit{the simple entropy}, and we prove that it holds under a set of assumptions.  The simple entropy is defined as a coarse graining of $S_{vN}$ obtained by maximizing $S_{vN}$  subject to fixing the expectation values of ``simple'' boundary operators with ``simple'' sources turned on.  We refer to sources defined on some set of boundary points $V \in \partial M$ as \emph{simple} if the bulk fields they produce propagates causally into the bulk from the points in $V$.  We will define a boundary operator \emph{simple} if the corresponding infinitesimal sources are simple.\footnote{The effects of finite sized sources are given by a time-ordered exponential, whereas in the case of operators we are only be interested in their expectation value; that is why we only require ``infinitesimal'' causality in the definition of simplicity for operators.  Thus the simple operators lie in a vector subspace of operators, while the set of simple sources may not have a vector space structure.}  

The reason why we call these operators ``simple'' is that sufficiently complicated operators in a region $R$, e.g. precursers \cite{PolSus99, FreGid02}, should be able to access data arbitrarily deep in the entanglement wedge of $R$, including in the inner wedge behind a minimar surface $\mu$.

In our classical bulk regime, all local operators are simple, and it should be sufficient to fix the one-point functions of these local operators (since the higher $n$-point functions are determined from these).  Furthermore it is sufficient to restrict attention to local simple sources, although not all local sources are simple, e.g. the exponentiation of the Hamiltonian $H$ can change the time-localization of fields acausally \cite{RobSta14}.

Our coarse-graining procedure to define the simple entropy is:
\begin{enumerate}
	\item[\textit{i.}] Choose a boundary initial time slice $t=t_{i}$, and a very late-time cutoff $t = t_f$ (in order to prevent recurrences),\footnote{We will assume that $t_f - t_i$ is much longer than any other time scale in the problem.} 
	\item[\textit{ii.}] Fix the one-point functions of local operators after $t_i$ (but before $t_f$) in the presence of all possible simple sources turned on after time $t_{i}$, but without changing the state $\rho$ at $t_i$ (so that there is retarded propagation from the sources). \label{2}
	\item[\textit{iii.}] Find the state $\rho'$ the maximizes the von Neumann entropy $S_{vN}$ over all of the states with the same simple one-point functions as defined in \ref{2} for $\rho$.
\end{enumerate}
In short: 
\be\label{eq:simple}
S^{\mathrm{(simple)}}[t_{i}]\equiv \max\limits_{\rho'} \left   [ 
S_{vN}[\rho']
\, : \,  
\langle E \mathcal{O} E^{\dagger} \rangle_{\rho'} = \langle E \mathcal{O} E^{\dagger} \rangle_\rho
\right ],
\ee
where 
\be\label{eq:E}
E = {\cal T}\, \text{Exp}\left[- i \int\limits_{t_{i}}^{t_f} J[t']{\cal O}_{J}[t']dt' \right],
\ee
$J[t']$ is a simple source, and ${\cal O}_{J}$ the corresponding simple operator.\footnote{As stated above, in the classical bulk regime, the ${\cal O}_{J}$'s can themselves be written as spatial integrals over local operators ${\cal O}_{J}(t',x')$, but for ease of notation we have not written the spatial dependence explicitly in Eq. \eqref{eq:E}.} (Note that $S^{\mathrm{(simple)}}[t_{i}]$ is not quite a purely boundary construct, as the definition of simple sources references the behavior of the corresponding bulk fields.  We hope that in the future, a purely boundary description of ``simple'' sources can be provided.)

\begin{figure}
	\begin{center}
		\includegraphics[width=0.5\textwidth]{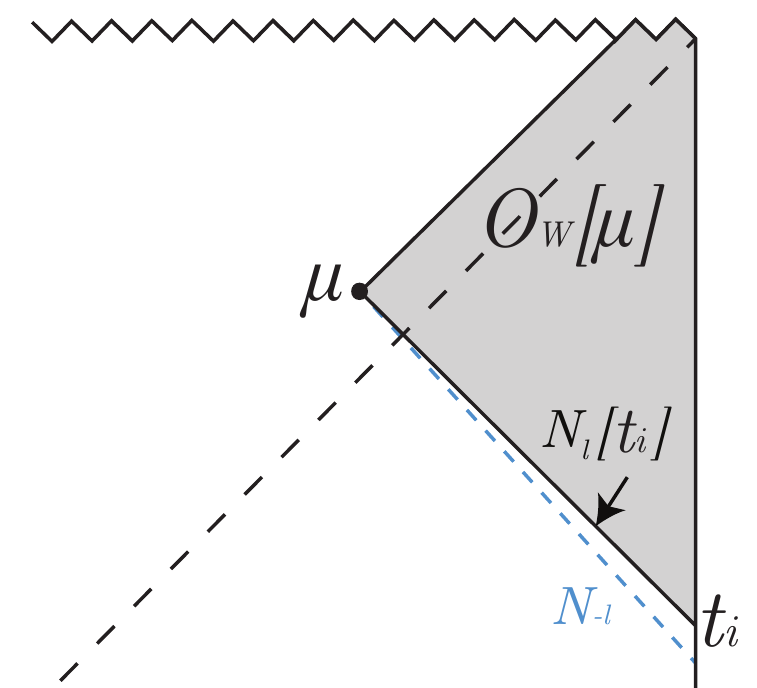}
	\end{center}
\caption{Generating a minimar surface by firing a null congruence $N_{\ell}[t_{i}]$ into the bulk in the $+\ell^{a}$ direction from $t_{i}$. This will not always coincide with $N_{-\ell}$ fired from $\mu$, but our procedure guarantees that there is no matter thrown into the bulk between $N_{-\ell}\cap \partial M$ and $t_{i}$.}
\label{fig:simple}
\end{figure}

We now wish to relate the simple entropy to the outer entropy of some minimar surface $\mu$ in the bulk.  The following construction is natural: take the slice $t_i$ and shoot in a future-directed null hypersurface $N_\ell[t_i] \equiv \partial I^+[t_i]$.  See Fig. \ref{fig:simple}.  In a black hole spacetime, there ought to exist some slice $\mu$ of $N_\ell[t_i]$ for which the outgoing expansion $\theta_{(k)}$ vanishes.\footnote{To prove this rigorously, we would need to analyze the analogue of Eq. \eqref{eq:loc} when $\theta_v \ne 0$.}  We expect that the outermost such slice $\mu$ generically satisfies $\nabla_\ell \theta_{(k)} < 0$;\footnote{On a \emph{smooth, spacelike} Cauchy slice, an outermost marginally trapped surface is generically guaranteed to exist \cite{AndMarMet08}, and satisfies the ``stability'' property $\nabla_\ell \theta{(k)} \le 0$ \cite{AndMar08}.  Since there always exist spacelike slices very close to $N_\ell[t_i]$, we therefore expect our strict form of stability to hold generically.} by focusing, $\mu$ has minimal area on the part of $N_\ell$ outside $\mu$.  Hence, $\mu$ should be minimar, at least generically.  Note that any such $\mu$ lies outside of any past horizon.

In this section, we consider only minimar surfaces $\mu$ that are obtained from boundary time slices $t_i$ in this manner.  There can exist other minimar surfaces which cannot be constructed in this way.  Note also that if $N_\ell[t_i]$ forms caustics before it reaches $\mu$, then $N_{-\ell}[\mu] $ (the past boundary of $O_{W}[\mu]$) will not coincide with $N_\ell[t_i]$!  But by bulk causality, $N_{-\ell}[\mu]$ lies nowhere to the future of $N_\ell[t_i]$.  However, in this case the domains of dependence agree: $N_{-\ell}[\mu] = N_{-\ell}[\mu]$.  This implies that the data in $O_W[\mu]$ can be reconstructed from the part of $O_W[\mu]$ which is to the future of $N_\ell[t_i]$, so long as we also know what the sources are between $t_\mu = N_{-\ell}[\mu] \cap \partial M$ and $t_i$.  See Fig. \ref{fig:simple}.

We now evaluate $S^{\mathrm{(simple)}}[t_{i}]$ by a three-step argument:  
\begin{enumerate}
\item First we hold the sources $J(t > t_i)$ fixed, and identify the ``reconstructible region'' $R[t_i]_J = D[I^+[t_i] \cap I^-[\partial M]]$ of the bulk which can be fully reconstructed from the one-point boundary data,\footnote{This is equivalent to the one-point data of \cite{KelWal13}, where their domain of dependence is taken to be the whole boundary.  That work erroneously conjectured that, in the case where there are no boundary sources, maximizing $S_{vN}$ subject to this one-point data would give Area$[N_\ell[t_i] \cap H^+]$, where $H^+ = \partial I^-[\partial M]$ is the future event horizon.} and that furthermore no bulk information independent of $R[t_i]_J$ can be recovered.  
\item Next we allow the sources to vary; in this case the size of $R[t_i]_J$ may depend on $J$, but it always remains within the outer wedge $O_W[\mu]$ of the new spacetime $M'$.  
Hence we can reconstruct at most $O_W[\mu]$, and it follows that $S^{\mathrm{(simple)}}[t_{i}] \ge S^{\mathrm{(outer)}}[\mu]$.  
\item Finally we wish to vary the sources $J(t > t_i)$ in such a way as to maximize the extent of $R[t_i]_J$.  We will argue that, for certain classes of states, there exists a $J$ such that $R[t_i]_J = O_W[\mu]$, so that \emph{all} of the data in $\partial^-O_W[\mu]$ is visible to the boundary, and hence
\be\label{SisO}
S^{\mathrm{(simple)}}[t_{i}] = S^{\mathrm{(outer)}}[\mu].
\ee
We can prove these results to all orders of perturbation theory around equilbrium (e.g. at late times in an AdS-Kerr ringdown process) --- and also of course for states that differ from such near-equilbrium states by the addition of large simple sources after time $t_i$.  We therefore conjecture that the equality in fact holds for \textit{all} classical states in the holographic regime.
\end{enumerate}

\paragraph{Step 1: $R[t_{i}]_{J}$ with fixed $J$:} Given the one-point data in a fixed state with fixed simple sources, we make use of the HKLL prescription~\cite{HamKab05, HamKab06, HamKab06b} to reconstruct $I^+[t_i] \cap I^-[\partial M]$, the region causally accessible from the boundary after $t_i$.  Using the bulk equations of motion, we can then reconstruct the domain of dependence of this region, which we call $R[t_{i}]_{J} = R[t_i]_J = D[I^+[t_i] \cap I^-[\partial M]]$.\footnote{In situations involving caustics, the domain of dependence is larger.}

The HKLL procedure solves a non-standard Cauchy problem by evolving the boundary one-point data ``sideways'' into the bulk via the bulk equations of motion; in this way the one-point functions can be used to reconstruct all local bulk operators\footnote{In a regime where gravitational back-reaction is important, it is necessary to ``dress'' these local bulk operators with suitable gravitational field lines extending out to the boundary $B$.} anywhere in $R[t_{i}]_{J}$. The HKLL procedure has been rigorously established for free field evolution at least if we assume any of (a) spherical symmetry \cite{HamKab06}, (b) Killing symmetry \cite{Tat07}, or (c) analytic bulk sources~\cite{Holmgren}.\footnote{There exists a counterexample to HKLL for a scalar field with evolution equation $\Box \phi = V \phi$, where $V$ is a complex potential (without analyticity, spherical symmetry, or stationarity) \cite{AliBao95}. However, this theory is unphysical due to $V$ being complex.} It is also possible to include interactions, at least perturbatively in 1/N \cite{Kab11,HeeMar}.  Below we will assume that the HKLL procedure can be performed perturbatively on general globally hyperbolic spacetimes, outside of event horizons.

Aside from the information in $R[t_i]_{J}$, no additional independent information about the spacetime $M$ can be reconstructed via the one-point data to the future of $t_{i}$. To see this, consider a modification of the fields localized in the spatially complementary region to $R[t_i]_{J}$, which we term $R^{c}[t_i]_J$. Since we are in a regime where bulk fields propagate causally, it is clear that such a modification cannot affect local operators on the boundary after the time $t_i$. Hence it is not encoded in our one-point data, and all of the reconstructible data is in $R[t_i]_J$.\footnote{This does not exclude the possibility that we may be able to deduce \emph{some} information outside of $R[t_i]_J$ from $R[t_i]_J$ using the bulk equations of motion and/or constraint equations, but since this information is determined by $R[t_i]_J$ it is not independent data.  That is why, in the argument above, we consider only valid initial data that does not change $R[t_i]_J$.}


\paragraph{Step 2: Upper bound on $R[t_{i}]_{J}$ with varying $J$:} 
We now allow arbitary simple boundary sources $J$ to be turned on after time $t_{i}$ (while holding fixed all sources before $t_{i}$).  Since the sources are simple, the resulting matter fields propagate causally into the bulk.  Hence the change in the geometry is localized in the region $I^+[t_{i}]$, and in particular the null hypersurface $N_\ell[t_i]$ is unaffected.  This allows us to compare the size of the invariant region $R[t_{i}]_{J}$ for two different choices of $J$, by comparing how much of the invariant hypersurface $N_\ell[t_i]$ is included in $R[t_{i}]_{J}$.

Turning on certain sources introduces more infalling matter and causes the event horizon to move outwards along $N_\ell[t_i]$; turning on other sources removes the infalling matter and push the event horizon inwards along $N_\ell[t_i]$.  However, there is no set of sources that can shift the event horizon so far inwards that $\mu$ lies outside of it~\cite{Haw71, EngWal14}. This is a consequence of a theorem of Hawking~\cite{Haw71}, which states that marginally trapped surfaces always lie behind event horizons. Thus, there is no set of simple sources $J'$ that we can turn on to produce a geometry in which $\mu$ lies inside $R[t_{i}]_{J'}$.  For the same reason, the null hypersurface $N_{+k}[\mu]$, which is foliated by trapped surfaces, also lies behind the event horizon.  And obviously $R[t_{i}]_{J'}$ also cannot dip to the past of $N_{-\ell}[\mu]$.  This shows that
\be
R(t_{i})_{J'}\subseteq O_{W}[\mu],
\ee
for any modified sources $J'$.

So far we have defined the reconstructible region $R[t_{i}]_{J'}$ on $M'$, the geometry corresponding to the modified sources $J'$.  Since the hypersurface $N_\ell[t_i]$ is invariant, we can use it to define the corresponding reconstructible region $R[t_{i}]_{J'}$ on $M$, the original manifold with sources $J$.  This is simply the domain of dependence of the part of $N_\ell[t_i]$ that we can reconstruct, which contains the same data on both spacetimes.  (This may well be larger than the region $R[t_{i}]_{J} \subset M$ which could have been reconstructed using the original sources, but in no case can it be larger than $O_{W}[\mu]$ since it still does not extend past $\mu$ on $N_\ell[t_i]$.)  In other words, to reconstruct a field $\phi$ somewhere in $R[t_{i}]_{J} \subset M$, we simply evolve it back to initial data on $N_\ell[t_i]$ using the \emph{original} boundary sources $J$, and then we turn on the \emph{new} sources $J'$ that move the causal horizon inwards, making it visible to the boundary after time $t_i$.

The one-point functions $\langle O\rangle_{J}$ thus allow us to fully reconstruct (via HKLL) \textit{at most} the outer wedge $O_{W}[\mu]$ of $M$. That is, the set of data used to compute the simple entropy is a (possibly improper) subset of the set of data used to compute the outer entropy. Since both entropies involve maximization subject to these constraints, we conclude that $S^{\mathrm{(simple)}}[t_{i}] \geq S^{\mathrm{(outer)}}[\mu]$, i.e. the simple entropy is either equivalent to, or else coarser than, the outer entropy.

\paragraph{Step 3: Maximizing $R[t_{i}]_{J}$:} We start by considering the case where $\rho$ is perturbatively close to a state $\rho_{stat}$ of thermal equilibrium, which is dual to a stationary geometry $M_{stat}$. In $M_{stat}$, $R(t_{i})=O_{W}[\mu]$ (because $\mu$ lies on the Killing horizon). In the perturbed state $\rho$, there exists matter falling across the event horizon $H_{EH}$, which we need to remove to cause $\mu$ to sit on $H_{EH}$.  To any order in perturbation theory, we can regard this matter as crossing $H_{EH}$ on the original background $M_{stat}$.  We modify the state by removing the matter crossing $H_{EH}$ while keeping the data on $N_{\ell}[t_i]$ fixed.\footnote{In order to comply with the No Hair Theorem~\cite{MTW, Isr67, Isr68, Car70}, it may be necessary to leave some matter crossing the horizon to the future of some very late time $t_f$, but this should make an exponentially small difference to the event horizon location at early times.  Note also that, since an equilbrium black hole should be stable under small perturbations, the perturbations to the fields should not diverge at late times.} This can be done in the bulk by attaching a stationary null hypersurface $N_{+k}[\mu]$ to $\mu$ via a similar construction as given in Sec.~\ref{sec:main}; we call this modified spacetime $M'$ (as in the previous step). We can now use HKLL to solve for the corresponding boundary sources $J'[t>t_i]$, which must differ from the original sources $J[t>t_i]$ since otherwise the spacetimes would be the same (and $\mu$ would already be a cross-section of $H_{EH}$). These are the sources that are needed to ``turn off'' matter falling across the horizon $H_{EH}$. Note that due to caustics and intersections, $N_{\ell}[t_{i}]$ need not coincide with $N_{-\ell}[\mu]$. This is fine, as $N_{-\ell}[\mu]$ will always lie to the past of $N_{\ell}[t_{i}]$, and thus no new sources are present in th region between the two congruences.

We now consider the boundary state which agrees with $\rho$ prior to $t_{i}$, but in which we turn on the $J'[t>t_i]$ rather than $J[t>t_i]$. Because $J'$ is simple, a classical bulk dual exists. This bulk dual is none other than $M'$, because the data on $N_{\ell}[t_i]$ together with the boundary sources $J[t>t_i]$ allows us to determine (via future-directed Cauchy evolution) the data on the horizon $H_{EH}$. This allows us to recover $O_{W}[\mu]$ to whichever order in perturbation theory we are working in. As explained in step 2, we can recover the outer wedge in $M$ as well as $M'$. Thus, we find that for $\rho$ perturbatively close to $\rho_{stat}$, the outer and simple entropies agree: $S^{\mathrm{(simple)}}[t_{i}] = S^{\mathrm{(outer)}}[\mu]$. 

To see why we only work perturbatively, consider the case where $\mu$ is a finite distance away from $H_{EH}$. In this case even if we turn off the matter on $H_{EH}$, we are not guaranteed that $\mu$ lies on $H_{EH}$, as there may be another minimar surface in the way. This is not possible in perturbation theory because on $N_{\ell}[t_i]$ there is a unique minimar surface on the Killing horizon $H_{EH}$ of the background spacetime $M_{stat}$, which remains unique under small perturbations.\footnote{This last point follows from the stability requirement that $\nabla_{k}\theta_{(\ell)}<0$ for a minimar surface.} Furthermore, it is possible that HKLL is valid only perturbatively, in which case we not justified in using it when $\mu$ is deep inside the black hole. 

It is also clear that this equality holds for a state $\rho'$ that is prepared from the near-equilibrium state $\rho$ above by turning on any additional simple sources after $t_i$, even if these sources are large (so that perturbation theory is not valid).

\section{Explanation for the Second Law}\label{sec:secondlaw}

One consequence of the equivalence between area and entropy for minimar surfaces is a statistical interpretation of the area law for certain sorts of local horizons~\cite{Hay93, AshKri02, BouEng15a}, as a second law of thermodynamics.

A hypersurface $\mathcal{H}$ foliated by marginally trapped surfaces (and satisfying certain regularity conditions) is known as a future holographic screen ~\cite{BouEng15a, BouEng15b} (or a future trapping horizon \cite{Hay93}).  They can be defined in a way which is local in time, but are highly nonunique. Such a hypersurface can be timelike, spacelike, or null in different parts of $\mathcal{H}$.  But if a black hole settles down to a stationary configuration, then at late times its causal horizon coincides with a null holographic screen $\mathcal{H}$.  These holographic screens obey an area law: the area of the marginally trapped surfaces increases with evolution along $\mathcal{H}$~\cite{Hay93, AshKri02,  GouJar06,BouEng15a,SanWei16} when moving towards the \emph{past} (on a timelike segment) or \emph{outwards} (on a spacelike segment).  

To apply our results, we need to consider the case where the holographic screen is foliated by minimar surfaces, which by Requirement \ref{def:minimarCross} satisfy $\nabla_\ell \theta_{(k)} < 0$.  From this inequality, it follows from the NCC that $\mathcal{H}$ must be spacelike or null~\cite{Hay93}, in which case they are also known as dynamical horizons \cite{AshKri02}.

In general there will be multiple holographic screens foliated by minimar surfaces on the same black hole background.  As an example, for any slicing of of a black hole spacetime $M$ into Cauchy hypersurfaces $\Sigma[t]$, the apparent horizon (i.e. the outermost marginally trapped surface on each $\Sigma(t)$) satisfies $\nabla_\ell \theta_{(k)} \le 0$ \cite{HawEll}, and thus --- assuming there is no homologous surface of lesser area outside of it\footnote{This ought to be true at least for small perturbations to a stationary black hole.} --- generically satisfies the criteria to be a minimar surface.  The evolution with $t$ would then define a holographic screen $\mathcal{H}$ foliated by minimar surfaces.  In general, the location of $\mathcal{H}$, and hence the outer wedges $O_W[\mu]$ used for coarse graining, will depend on the choice of foliation $\Sigma[t]$.  Our derivation of the second law will hold for \emph{any} holographic screen $\mathcal{H}$ foliated by minimar surfaces, whether or not it is obtained by one of the construction described in this paragraph.

As stated above, the area of the minimar surfaces $\mu$ will increase as we evolve outwards along $\mathcal{H}$:
\be\label{areainc}
\text{Area}[\mu_2] \ge \text{Area}[\mu_1],
\ee
where $\mu_2$ is further outwards than $\mu_1$.  This can be proven by geometrical means using the NCC, but we will now provide an simple statistical-mechanical explanation for the area increase in terms of the outer entropy.

\begin{figure}
	\centering
	\includegraphics[width=0.33\textwidth]{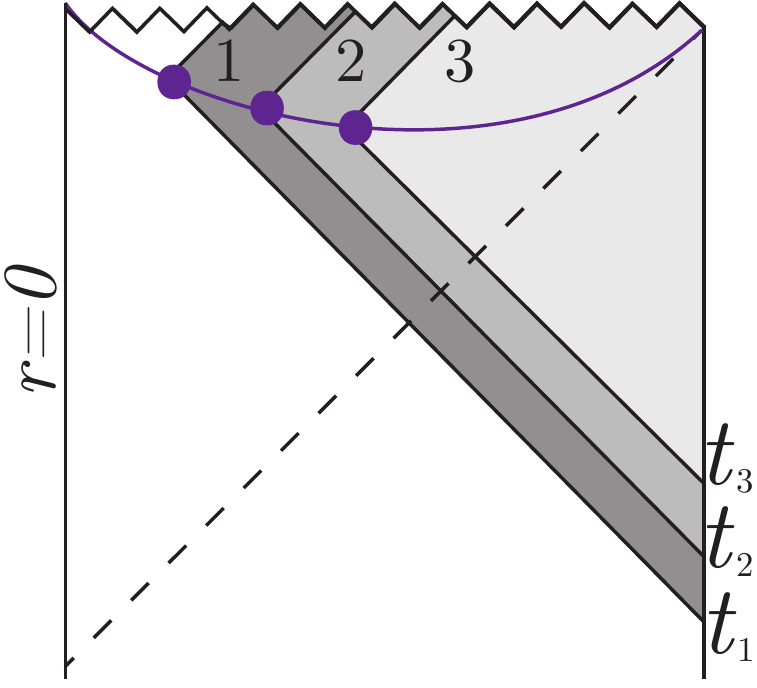}
	\caption{Reproduced from~\cite{EngWal17b}. The outer wedges of the marginally trapped surfaces constituting a holographic screen are nested: evolution along the screen from the leaves labeled 1, 2, and 3 corresponds to progressively larger outer entropy. On the boundary, this translates into a timelike law of outer entropy increase: $S^{\mathrm{(outer)}}[t_{1}]<S^{\mathrm{(outer)}}[t_{2}]<S^{\mathrm{(outer)}}[t_{3}]$.}
	\label{fig:nesting}
\end{figure}

First note that the corresponding outer wedges are nested: $O_W[\mu_2] \subset O_W[\mu_1]$.  Hence there is less data in $O_W[\mu_2]$ than $O_W[\mu_1]$.  This is illustrated in Fig.~\ref{fig:nesting}. It follows that 
\be
S^{\text{(outer)}}[\mu_2] \ge S^{\text{(outer)}}[\mu_1], 
\ee
since we are maximizing $S_{vN}$ with respect to fewer constraints at $\mu_2$.  But $S^{\text{(outer)}}[\mu] = \text{Area}[\mu]$ for each surface, so the area increase inequality \eqref{areainc} follows automatically.\footnote{Unfortunately, we do not know how to give an explanation of the area law for the timelike (or mixed signature) parts of holographic screens.  In this case, the marginally trapped surfaces on the timelike component have $\nabla_{k}\theta_{(\ell)}>0$ and are thus not minimar surfaces, so we cannot prove that their area corresponds to the entropy associated to ignorance of their interior.  Another obstacle is that in the timelike case, their outer wedges $O_W[\mu]$ are not nested: we cannot yet explain thermodynamically why the entropy on the timelike component of the holographic screen increases \textit{towards the past}. This apparent contradiction with our usual intuitions about the direction of entropy increase remains a mystery.}

This second law also has an appealing interpretation on the boundary side.  Suppose that we obtain our holographic screen $\mathcal{H}$ shooting in null surfaces $N_\ell(t)$ from a Cauchy foliation $\Sigma[t]$ of the \emph{boundary} $\partial M$, as in Sec. \ref{sec:simple}.  (Such a holographic screen always lies outside of any past horizons.)  In this case, the boundary interpretation of the growth of $\mathcal{H}$ is an increase in the simple entropy $S^{\text{(simple)}}[t_i = t]$.  (Here we are holding the very late time cutoff $t_f$ fixed as we vary $t_i$).  Consider two different initial times $t_i = t_1$ and $t_i = t_2$, with $t_2 > t_1$.  Because we can only turn on simple sources and measure simple operators after $t_i$, it follows that there are fewer simple experiments that can be done after $t_2$ than after $t_1$.  Hence, 
\be
S^{\text{(simple)}}[t_2] \ge S^{\text{(simple)}}[t_1].
\ee
since the later entropy is maximized subject to fewer constraints.  Since at any time $S^{\text{(simple)}}[t_i] = S^{\text{(outer)}}[\mu[t_i]]$, this second law is equivalent to the previous ones, but now it is expressed in terms of boundary quantities.

Our construction suggests a natural perspective on proving the ordinary second law of theormodynamics even in non-holographic theories.  The reader may think it odd that $S^{\text{(simple)}}[t_i]$ has been defined relative to a set of measurements taking place at \emph{all} times later than $t_i$ (up to some very late cutoff $t_f$), rather than being restricted to times near $t_i$.  But this is actually very natural from the perspective of proving the second law.

First recall the standard (not completely satisfactory) textbook analysis of the second law.  Suppose for example we start with a pure state at time $t_0$ and then allow it to begin to thermalize.  To define a nontrivial second law (where entropy increases with time), we need a notion of coarse-graining such which allows us to ``forget'' some information that was available at $t_0$, once we have arrived at a later time $t_1$ when this information is no longer accessible to macroscropic measurements.  This allows us to define a coarse-grained entropy $S^{\text{(coarse)}}[t_1] > 0$.  However, if the forgotten information has not fully thermalized, then there is the danger that at a still later time $t_2$, the forgotten information may re-emerge into the macroscopic degrees of freedom, causing a decrease of $S^{\text{(coarse)}}$ from $t_1$ to $t_2$!  It is very hard to prove rigorously that this cannot happen in reasonable matter systems.

Our approach neatly sidesteps this issue by defining the coarse-grained entropy relative to observations made \emph{anywhere} in the time interval $(t_i, t_f)$, with the late time cutoff $t_f$ taken very large.  That is equivalent to saying, that if any reasonable future experiment could have recovered some piece of information, then (almost by definition) we ought \emph{not} to have coarse-grained over that piece of information for purposes of the second law, since it has not been irreversibly thermalized into the microscopic degrees of freedom.  Maximizing $S_{vN}$ subject to to all information accessible in $(t_i, t_f)$ automatically excludes such pathological cases of information return, and makes it easy to prove a second law mathematically for all systems, without the use of additional postulates that are difficult to justify.  

The price that we pay is that such an increasing coarse-grained entropy may be hard to evaluate, in situations where it is unclear whether some information is permanently lost.  Fortunately, this turns out not to be an issue in the holographic context, since the duality to black hole horizons makes a sharp division between the information that is accessible, and the information that is (classically) lost forever.

\section{Prospects}\label{sec:prospects}

We have shown that in black hole physics, the area of certain marginally trapped ``minimar surfaces'' have a natural interpretation as a coarse-grained holographic entropy, which we have called the ``outer entropy''.  We have given a statistical explanation for why the corresponding holographic screens obey a second law, and (at least perturbatively) have shown that this is equivalent to a second law on the boundary, expressed in terms of the ``simple entropy''.  As described at the end of the previous section, this boundary second law provides an interesting new perspective on the thermodynamics of ordinary (not necessarily holographic) systems.

Leaving aside our proposed boundary dual, our only use of holography in the bulk has been the HRT formula for the holographic entanglement entropy \eqref{HRTintro}.  Thus, all of our bulk results about the outer entropy also extend to the case of asymptotically flat bulk black holes, assuming that (as is plausible) the area of extremal surfaces also corresponds to some von Neumann entropy in this case (perhaps, defined in terms of a hypothetical flat space holographic dual -- see \cite{BarCom06, Bag10, BagDet12, Com12, BagBas14, JiaSon17} and references therein).  

Another natural extension of our work is to boundary-anchored marginally trapped surfaces in AdS.  We expect that similar results will hold, but in this article we have only covered the case of nonminimal extremal surfaces (see Sec. \ref{sec:extremal} for the details).

This article has not explained the second laws that are known to be obeyed by timelike holographic screens \cite{BouEng15a, BouEng15b} and by causal horizons \cite{Haw71}.  Although holographic screens obey a second law in cosmology \cite{BouEng15c}, it is especially unclear how to extend our results to this case, since in a closed universe the minimal area surface dual to e.g. a de Sitter horizon is always the empty set.

Another direction that remains to be addressed is the extension to semiclassical settings, in which the black hole is coupled to quantum field theory.  In this case, we expect that we need to replace the area with a \emph{generalized entropy} which includes bulk entropy corrections \cite{FauLew13, EngWal14}.  Hence, we will need to consider \emph{quantum} marginally trapped surfaces \cite{EngWal14}, and we will end up with a second law for certain \emph{Q-screens} \cite{BouEng15c}.  However, in order to construct our stationary null surface $N_{-k}$, we will need a better understanding of when, given the data outside the surface, we can saturate inequalities such as monotonicity of relative entropy.  See \cite{Wal17PRL} for discussion of a relevant conjecture.

If our results can be extended to the semiclassical regime, they are likely to provide an interesting perspective on the firewalls puzzle \cite{AMPS, AMPSS, Mat09}.  Recall that the paradox here is that strong subadditivity seems to prevent old black holes (that are highly entangled with their early Hawking radiation) from having a normal interior.  A quantum version of our result could be used to construct the ``best possible'' (i.e. entropy maximizing) interior of the black hole as a function of time, which might reveal interesting behavior across the transition to the ``firewall'' phase.

Finally, we would like to speculate on what our results mean for nonperturbative quantum gravity.  It is natural to suppose that the Bekenstein-Hawking entropy of any surface $\sigma$ corresponds to the entropy of some set of Planck-scale boundary qubits sitting on $\sigma$ \cite{Sor83, Jac95, Sor05, BiaMye12}.  If these qubits can be approximately localized, this explains why the entropy is an extensive (geometric) intergral on $\sigma$.  

Our results show that if $\sigma$ is a minimar surface, it is possible to act on the state in a way that maximally mixes the qubits, without changing the classical geometry outside of $\sigma$.  These degrees of freedom can therefore be regarded as independent degrees of freedom.  On the other hand, for an untrapped surface, it is \emph{not} usually possible to fully mix the qubits without changing the geometry outside (see Sec. \ref{sec:untrapped}).  So these degrees of freedom cannot become fully mixed without adding energy from outside.  Finally, in the case of a trapped surface, the outer entropy can exceed the total entropy of the surface qubits.  In this case, there must be some other source of boundary entropy which is not fully accounted for by the Planckian qubits near $\sigma$.

For a model of holographic quantum gravity in the bulk to be successful, it must be able to explain why there is a match between the area and the outer entropy for minimar surfaces, but not for these other classes of surfaces.

\section*{Acknowledgments}
It is a pleasure to thank S. Alexakis, R. Bousso, S. Fischetti, G. Horowitz, H. Kunduri, J. Maldacena, D. Marolf, F. Pretorius, G. Remmen, A. Shao, D. Stanford, H. Verlinde, S. Weinberg, E. Witten, and B. White for discussions.  The work of NE was supported in part by NSF grant PHY-1620059 and in part by the Simons Foundation, Grant 511167 (SSG). NE thanks the Stanford Institute for Theoretical Physics for hospitality during the final stages of this work. The work of AW was supported in part by NSF grant PHY-1314311, the Stanford Institute for Theoretical Physics, the Simons Foundation (``It from Qubit''), the Institute for Advanced Study, and the S. Raymond and Beverly Sackler Foundation Fund.

\appendix

\section{HRT Surfaces are Minimar Surfaces}\label{sec:HRT}

We prove in this section that HRT surfaces are automatically minimar.  

By definition, HRT surfaces are homologous to the boundary; by the maximin method~\cite{Wal12}, they are also minimal on a complete Cauchy slice.  Hence they satisfy Requirement~\ref{def:minimarMin} for a minimar surface.  To show Requirement~\ref{def:minimarCross}, that $\nabla_{k}\theta_{(\ell)} \le 0$ we prove the following:

\paragraph{Theorem:} Let $X$ be an extremal surface homologous to $B$ which is minimal on a Cauchy slice $\Sigma$ of $O_{W}[X]$. Then $\nabla_{k}\theta_{(\ell)} \le 0$. 
\begin{proof}
Consider firing out the null congruence $N_{+k}[X]$ in the $k$ direction from $X$, and let $\sigma$ be a cross-section of $N_{+k}[X]$. We may now fire the null congruence $N_{-\ell}(\sigma)$ from $\sigma$ in the $-\ell$ direction (i.e. towards the past and away from $X$), towards the slice $\Sigma$ on which $X$ is the minimal area surface. We know that $\sigma'=N_{-\ell}(\sigma)\cap \Sigma$ must have area greater than or equal to the area of $X$. By taking $\sigma$ to be sufficiently close to $X$, we can guarantee that $\sigma'\subset U$, where $U$ is any open set on $\Sigma$. Then:
\be
\mathrm{Area}[\sigma']\geq \mathrm{Area}[X]\geq\mathrm{Area}[\sigma]
\ee
where the first inequality follows from the minimality of $X$ and the second inequality follows by the focusing theorem. This means that the area of cross-sections of $N_{-\ell}(\sigma)$ has to grow (or remain unchanged) in moving from $\sigma$ to $\sigma'$, i.e. the expansion $\theta_{(\ell)}$ at some point on $N_{-\ell}(\sigma)$ between $\sigma$ and $\sigma'$ has to be nonpositive. By the null energy condition, once the expansion is negative, it remains nonpositive. This means that $\left. \theta[N_{-\ell}(\sigma)]\right|_{\sigma}\leq0$. 

Since $\sigma$ may be taken to be arbitrarily close to $X$, we find:
\be
\theta_{\ell,k}[X]\leq 0.
\ee

\end{proof}

\bibliographystyle{JHEP}

\bibliography{all}

\end{document}